\documentclass[final,3p,sort&compress]{elsarticle}

\usepackage{amsmath}
\usepackage{amssymb}
\usepackage[figurename=Fig.,labelsep=period]{caption}
\usepackage[colorlinks,citecolor=blue,urlcolor=blue,linkcolor=blue]{hyperref}

\begin{document}

\begin{frontmatter}



\title{Scattering of fermions on a one-dimensional Q-ball}

\author[tusur]{A.Yu.~Loginov}
\ead{a.yu.loginov@tusur.ru}

\address[tusur]{Tomsk State University of Control Systems and Radioelectronics, 634050 Tomsk, Russia}

\begin{abstract}
The scattering  of  massless  fermions  on  a one-dimensional Q-ball is studied
both analytically and numerically in the background field approximation.
The wave functions of  the  fermionic scattering states are found in analytical
form.
General expressions are derived for the transmission and reflection coefficients
and the corresponding $S$-matrix elements.
General formulae  describing  the  evaporation  of  the  Noether  charge of the
one-dimensional Q-ball are given.
A numerical study of the transmission  and  reflection  coefficients along with
the corresponding $S$-matrix elements is performed for a range of values of the
model  parameters.
A study of the dependence of the  evaporation  rate of the Q-ball on the Yukawa
coupling constant is carried  out for several values of the Noether charge.
\end{abstract}


\end{frontmatter}

\section{Introduction} \label{sec:I}

Many field models possessing an unbroken global symmetry admit the existence of
nontopological solitons \cite{lee_pang_1992, radu_volkov_2008}.
Nontopological solitons  are  spatially  localized finite-energy solutions with
topologically trivial field configurations.
Unlike topological solitons \cite{Manton, Shnir}, which  possess  a  nontrivial
topology, topological triviality cannot ensure  the stability of nontopological
solitons.
The main property of  a  nontopological  soliton is that it is a local extremum
(minimum or saddle point) of the  energy  functional  for  a fixed value of the
conserved Noether charge.
Under certain conditions, this  extremum  is  an absolute minimum of the energy
functional and the nontopological soliton  is the ground state in the sector of
a fixed Noether charge.
In this case, the stability of the nontopological soliton is due to conservation
of the Noether charge.

The  simplest  and   most  important   nontopological    soliton,  proposed  in
Ref.~\cite{rosen_1968_a}  and known as the Q-ball \cite{coleman_1985}, has been
found in a $U(1)$-invariant model  of a self-interacting complex  scalar field.
In Refs.~\cite{saf1, saf2}, it was shown that Q-balls can  also exist in scalar
field models possessing global non-Abelian symmetry.
Furthermore, $U(1)$  gauged  models  of  complex self-interacting scalar fields
also admit the existence of Q-balls \cite{rosen_1968_b,  klee,  lee_yoon, anag,
 levi, ardoz_2009, benci, tamaki_2014,  gulamov_2014,  brihaye_2014, hong_2015,
 gulamov_2015, loginov_prd_102}.
In realistic models, Q-balls are generally allowed in supersymmetric extensions
of the Standard Model that have flat  directions  in  the interaction potential
of scalar fields \cite{kus_plb_1997_405, kst_1998, enqmcd_1998}.
These Q-balls are of  great  interest  to  cosmological  models  describing the
evolution of the early Universe \cite{dksh_plb_417, kusshp_plb_418, enqmcd_1999,
kk_2000, enqmzm_2003, kky_2013a, kky_2013b, cotner_2017a, cotner_2017b}.
In some models, Q-balls can survive to  the  present as places of concentration
of  dark  matter  \cite{kus_plb_1997_405, kusshp_plb_418},  whereas  in  other,
Q-balls decay and do not survive to the present.
In the latter case, Q-ball decay may result in the production of dark matter in
the form of the lightest supersymmetric particles \cite{enqmcd_1999}.

In realistic  models,  scalar  fields  forming  a  Q-ball interact with fermion
fields.
This interaction may  have  important  consequences  for  the  stability of the
Q-ball \cite{ccgm, kvng}.
In particular,  it  was  shown  in  Ref.~\cite{ccgm}  that  the  interaction of
massless fermions with a scalar field leads to evaporation of the Q-ball.
A numerical study of  this process, taking into account real profiles of Q-ball
solutions, was performed in Ref.~\cite{mv}.
A detailed study of the evaporation of  a  one-dimensional Q-ball was performed
in Ref.~\cite{clark}, where the real profile of the Q-ball was approximated  by
a rectangular one.

Evaporation of the Q-ball is possible only within a limited range of the energy
parameter.
At larger values of the energy parameter, fermion-Q-ball scattering takes place
rather than fermionic evaporation of the Q-ball.
In the present work, we  investigate  the  scattering of massless fermions on a
one-dimensional Q-ball.
We also investigate the fermionic evaporation of the Q-ball within an allowable
region of the energy parameter.
The choice of  a  one-dimensional  Q-ball  was  made  due  to  the fact that an
analytical solution is known only for this case \cite{lee_pang_1992}.
This makes it possible to obtain  analytical expressions for the fermionic wave
functions in the background field of the one-dimensional Q-ball.
In turn, the analytical expressions of the fermionic wave functions allow us to
obtain general  expressions  for  the  fermionic  transmission  and  reflection
coefficients,  which  considerably  facilitates  the   study  of fermion-Q-ball
scattering.

This paper is structured as follows.
In  Sec.~\ref{sec:II},  we   give  a  concise  description  of  the Lagrangian,
symmetries, and field equations of the model, and give the analytical form of a
one-dimensional Q-ball solution.
Section~\ref{sec:III} presents  an  analytical  description  of  fermion-Q-ball
scattering.
In particular, analytical  expressions  for  the  fermionic  wave functions and
general  expressions  for  the  transmission  and  reflection  coefficients are
presented.
In Sec.~\ref{sec:IV}, fermionic  evaporation  of  the Q-ball is considered, and
formulae describing this process are given.
Section~\ref{sec:V} contains numerical results.
In particular, we  discuss  the  dependence  of  the  fermion  transmission and
reflection coefficients on an energy parameter.
We also discuss the  dependence  of  the  evaporation rate of the Q-ball on the
value of the Yukawa coupling constant.
In the final section,  we  briefly summarise the results obtained in this work.

Throughout the paper, we use the natural units $\hbar = c = 1$.

\section{Lagrangian and field equations of the model}            \label{sec:II}

The bosonic part of the model we are interested  in  has the Lagrangian density
\begin{equation}
\mathcal{L}_{\text{b}}=\partial _{\mu }\phi \partial^{\mu}\phi^{\ast}-V\left(
\left\vert \phi \right\vert \right),                               \label{II:1}
\end{equation}
where
\begin{equation}
V\left( \left\vert \phi \right\vert \right) =m^{2}\left\vert \phi
\right\vert ^{2}-\frac{g}{2}\left\vert \phi \right\vert ^{4}+\frac{h}{3}
\left\vert \phi \right\vert^{6}                                    \label{II:2}
\end{equation}
is the self-interaction potential of the complex scalar field $\phi$.
In Eq.~(\ref{II:2}), the  coupling  constants  $g$  and  $h$  are assumed to be
positive and to satisfy the inequality $m^{2} h g^{-2} > 3/16$.
It follows that the absolute minimum of the  potential $V\left( \left\vert \phi
\right\vert \right)$ occurs at $\phi = 0$,  and  the  potential vanishes there.

The Lagrangian density  (\ref{II:1})  is  invariant  under  the  global $U(1)$
transformations
\begin{equation}
\phi \left( t, x\right) \rightarrow \phi^{\prime }\left(t, x\right) = \exp
\left(-i\alpha \right) \phi \left(t, x\right).                     \label{II:3}
\end{equation}
We want to introduce fermions in such  a  way that  the resulting model remains
invariant under transformations (\ref{II:3}).
We also want the resulting model to  have a conserved fermion current.
To do this, we rewrite Eq.~(\ref{II:1}) in terms of the real and imaginary parts
of the complex scalar field $\phi = 2^{-1/2}\left(\phi_{1} + i\phi_{2}\right)$,
where $\phi_{1}$  and  $\phi_{2}$  are  regarded  as  components  of the scalar
isotriplet $\boldsymbol{\phi} = \left(\phi_{1}, \phi_{2}, 0\right)$ with a zero
third component.
We then introduce a fermionic  isodublet $\psi$ that interacts  with the scalar
isotriplet  $\boldsymbol{\phi}$   via  the  Yukawa  interaction  to  obtain the
Lagrangian density
\begin{equation}
\mathcal{L} =\frac{1}{2}\partial _{\mu }\boldsymbol{\phi \cdot }\!\partial
^{\mu}\boldsymbol{\phi}-\frac{m^{2}}{2}\boldsymbol{\phi\!\cdot\!\phi}+\frac{
g}{8}\left( \boldsymbol{\phi \!\cdot \!\phi }\right) ^{2}-\frac{h}{24}\left(
\boldsymbol{\phi \!\cdot \!\phi }\right)^{3}
+i\bar{\psi}\gamma^{\mu }\partial _{\mu}\psi - G\boldsymbol{\phi
\cdot}\!\bar{\psi }\boldsymbol{\tau }_{\perp }\psi,                \label{II:4}
\end{equation}
where $G$ is the Yukawa coupling  constant, $\boldsymbol{\tau}_{\perp} = \left(
\tau_{1}, \tau _{2} \right)$, and $\tau_{1,\,2}$  are  the  corresponding Pauli
matrices.
In Eq.~(\ref{II:4}), the two indices of the Dirac field $\psi_{i a}$ correspond
to its spin-isospin structure.
We use the following set of Dirac matrices in $(1+1)$ dimensions:
\begin{equation}
\gamma^{0} =   \sigma_{1},\;
\gamma^{1} = -i\sigma_{2},\;
\gamma_{5} =   \gamma^{0}\gamma^{1}=\sigma_{3},                    \label{II:5}
\end{equation}
where $\sigma_{k}$ are the Pauli matrices.
To distinguish the Pauli  matrices  $\sigma_{k}$ acting on the spinor index $i$
of the fermionic field $\psi_{i a}$ from  those  acting  on its isospinor index
$a$, we denote the latter as $\tau_{k}$.

The Lagrangian (\ref{II:4}) depends on  the  four  dimensional parameters: $m$,
$g$, $h$, and $G$.
By scaling  the  space-time  coordinates,  fields,  and  coupling  constants as
\begin{equation}
x^{\mu } \rightarrow m^{-1}x^{\mu },\;
\boldsymbol{\phi }\rightarrow mg^{-1/2}\boldsymbol{\phi},\;
\psi \rightarrow m^{3/2}g^{-1/2}\psi,\;
h \rightarrow m^{-2}g^{2}h,\;
G\rightarrow g^{1/2}G,                                            \label{II:5a}
\end{equation}
the number of parameters of  the Lagrangian (\ref{II:4}) can be reduced to two.
After scaling  of  Eq.~(\ref{II:5a}),  the  Lagrangian (\ref{II:4}) transforms
as $\mathcal{L}  \rightarrow  m^{4}  g^{-1}  \mathcal{L}_{\text{s}}$, where the
scaled Lagrangian
\begin{equation}
\mathcal{L}_{\text{s}} = \frac{1}{2}\partial _{\mu }\boldsymbol{\phi \cdot }\!
\partial^{\mu}\boldsymbol{\phi}-\frac{1}{2}\boldsymbol{\phi\!\cdot\!\phi}+
\frac{1}{8}\left( \boldsymbol{\phi \!\cdot \!\phi }\right) ^{2}-\frac{h}{24}
\left(\boldsymbol{\phi \!\cdot \!\phi }\right)^{3}
+i\bar{\psi}\gamma^{\mu }\partial _{\mu}\psi - G\boldsymbol{\phi
\cdot}\!\bar{\psi }\boldsymbol{\tau }_{\perp }\psi                \label{II:4s}
\end{equation}
depends only on the two parameters $h$ and $G$.
Note that   the   scaled   fields    and    coupling   constants  included  in
Eq.~(\ref{II:4s})  are  dimensionless,  as  are  the space-time coordinates on
which these fields depend.

The Lagrangian (\ref{II:4s}) is invariant under the global transformations
\begin{subequations}                                               \label{II:6}
\begin{eqnarray}
\boldsymbol{\phi} &\rightarrow & \boldsymbol{\phi}^{\prime} =
\exp \left(-i \alpha T_{3}\right) \boldsymbol{\phi},              \label{II:6a}
\\
\psi  &\rightarrow &\psi ^{\prime } =
\exp \left( -i\alpha t_{3} \right) \psi,                          \label{II:6b}
\end{eqnarray}
\end{subequations}
where the generators $T_{3} = -i\epsilon_{3 a b}$ and $t_{3} = \tau_{3}/2$.
These transformations form  the  Abelian  subgroup of the corresponding $SU(2)$
group,  and   hence   Eqs.~(\ref{II:3})   and   (\ref{II:6a})  are  isomorphic.
The invariance of the  Lagrangian  (\ref{II:4s}) under  global  transformations
(\ref{II:6}) results in the conserved Noether current
\begin{equation}
j_{3}^{\mu}=-\epsilon_{3ab}\left(\partial^{\mu}\phi_{a}\right)\phi_{b}
+\frac{1}{2}\bar{\psi }\gamma ^{\mu }\otimes \tau _{3}\psi.        \label{II:7}
\end{equation}
In addition to transformations  (\ref{II:6}), the  Lagrangian  (\ref{II:4s}) is
also invariant under the global $U(1)$ transformations
\begin{equation}
\psi \rightarrow \psi^{\prime} = \exp\left(-i \beta \right)\psi,   \label{II:8}
\end{equation}
resulting in the conserved fermion current
\begin{equation}
j_{F}^{\mu} = \bar{\psi}\gamma^{\mu } \otimes \mathbb{I} \psi.     \label{II:9}
\end{equation}

The variation of  the  action $S = \int \mathcal{L} dx dt$ in the corresponding
fields leads us to the field equations:
\begin{equation}
\partial _{\mu }\partial ^{\mu }\boldsymbol{\phi }+\boldsymbol{\phi }-\frac{1
}{2}\left( \boldsymbol{\phi \! \cdot \! \phi }\right) \boldsymbol{\phi }+
\frac{h}{4}\left( \boldsymbol{\phi \! \cdot \! \phi }\right) ^{2}
\boldsymbol{\phi} + G\bar{\psi}\mathbf{\tau}_{\perp}\psi = 0     \label{II:10a}
\end{equation}
and
\begin{equation}
i\gamma^{\mu}\partial_{\mu}\psi - G \boldsymbol{\phi \cdot \!}
\boldsymbol{\tau}_{\perp }\psi = 0.                              \label{II:10b}
\end{equation}
It is known  that  under  the  condition  $h > 3/16$, model (\ref{II:4s}) has a
non-topological soliton solution called a Q-ball \cite{coleman_1985}.
In the $(1 + 1)$-dimensional case, the  Q-ball solution  can  be written in the
analytical form:
\begin{subequations}                                              \label{II:11}
\begin{eqnarray}
\phi _{1}\left( t,x\right)  &=&2\sqrt{2}\Omega \left( 1+\kappa \cosh \left(
2\Omega x\right)\right)^{-\frac{1}{2}}\cos\left(\omega t\right), \label{II:11a}
\\
\phi _{2}\left( t,x\right)  &=&-2\sqrt{2}\Omega \left( 1+\kappa \cosh \left(
2\Omega x\right)\right)^{-\frac{1}{2}}\sin\left(\omega t\right), \label{II:11b}
\\
\phi _{3}\left( t,x\right)  &=&0,                                \label{II:11c}
\end{eqnarray}
\end{subequations}
where $\Omega = \left(1 - \omega^{2}\right)^{1/2}$ and $\kappa = \left(1-\left(
16/3\right) h \Omega^{2}\right)^{1/2}$.
The parameter $\omega$ on  which  solution  (\ref{II:11})  depends is the phase
frequency of the complex scalar field $\phi = 2^{-1/2}(\phi_{1} + i \phi_{2})$.
The phase frequency  of the Q-ball solution satisfies the condition $\left\vert
\omega \right\vert  \in \left( \omega _{\text{tn}}, 1 \right)$, where $\omega_{
\text{tn}} = \left(1 - 3/\left(16 h\right) \right)^{1/2}$.
The main property  of  Q-ball  solution (\ref{II:11}) is that it is an absolute
minimum of the energy functional in the sector of  scalar  field configurations
with a fixed Noether charge $Q = \int j_{3}^{0}dx$.
It follows that Q-ball  solution (\ref{II:11}) is  stable over the entire range
of the parameter $\omega$.
As $\left\vert\omega\right\vert\rightarrow 1$, the energy and Noether charge of
the one-dimensional Q-ball solution (\ref{II:11}) tend to zero as $E \sim \left
\vert Q\right\vert\propto \left(1-\left\vert \omega \right\vert \right)^{1/2}$.
In contrast, as $\left\vert\omega\right\vert\rightarrow\omega_{\text{tn}}$, the
energy and  Noether  charge  of  the  Q-ball diverge logarithmically as $E \sim
\omega_{\text{tn}} \left\vert Q\right\vert \propto - \ln\left(\left\vert \omega
\right\vert - \omega_{\text{tn}}\right)$.

The Q-ball solution in Eq.~(\ref{II:11}) has the following  symmetry properties
under space and time reflections:
\begin{eqnarray}
\phi _{i}\left(t, -x\right)  &=&\phi _{i}\left(t, x\right),      \label{II:12a}
 \\
\phi_{i}\left(-t, x\right)  &=&\left(-1\right)^{i + 1}\phi_{i}
\left(t,x\right),                                                \label{II:12b}
\end{eqnarray}
and, as a consequence of Eq.~(\ref{II:12b}),
\begin{equation}
\boldsymbol{\phi}\left(t,x\right)\!\boldsymbol{\cdot}\!\boldsymbol{\tau}_{\perp
}=\boldsymbol{\phi}\left(-t,x\right)\!\boldsymbol{\cdot }\!\boldsymbol{\tau}
_{\perp }^{\ast}.                                                \label{II:12c}
\end{equation}
Using  symmetry  properties (\ref{II:12a})--(\ref{II:12c}),  it  can  easily be
shown that if $\psi(t, x)$ is a solution  to  the Dirac equation (\ref{II:10b})
in the background field of Q-ball (\ref{II:11}), then
\begin{subequations}                                              \label{II:13}
\begin{eqnarray}
\psi ^{C}\left( t,x\right)  &=&\eta _{C}\gamma _{5}\otimes \tau _{1}\psi
^{\ast }\left( t,x\right),                                       \label{II:13a}
 \\
\psi ^{P}\left( t,x\right)  &=&\eta _{P}\gamma ^{0}\otimes \mathbb{I}\psi
\left( t,-x\right),                                              \label{II:13b}
 \\
\psi ^{T}\left( t,x\right)  &=&\eta _{T}\gamma ^{0}\otimes \mathbb{I}\psi
^{\ast }\left( -t,x\right)                                      \label{II:13c}
\end{eqnarray}
\end{subequations}
are also solutions to the Dirac equation (\ref{II:10b}) in the background field
of the Q-ball, where $\eta_{C}$, $\eta_{P}$, and $\eta_{T}$  are phase factors.

Under the  reflection $\omega \rightarrow - \omega$,  the energy $E\left(\omega
\right)$  and   Noether  charge  $Q\left( \omega \right)$  of  Q-ball  solution
(\ref{II:11}) are even and odd functions of $\omega$, respectively.
Using Eq.~(\ref{II:12c}), it can be shown that if $\psi\left(t,x,\omega\right)$
is a solution to the Dirac  equation  in  the  background  field  of the Q-ball
corresponding to the phase frequency $\omega$, then
\begin{equation}
\psi \left( t,x,-\omega \right) =\mathbb{I}\otimes \tau _{1}\psi \left(
t,x,\omega \right)                                                \label{II:14}
\end{equation}
is a solution to the Dirac  equation  in  the  background  field  of the Q-ball
corresponding to the phase frequency $-\omega$.
It follows that we can limit ourselves  to  studying the case of positive phase
frequencies.

\section{Scattering of fermions in the background field of the Q-ball}
                                                                \label{sec:III}
To obtain analytical expressions for the fermionic  wave functions, we consider
fermion-Q-ball  scattering  in  the background field approximation, in which we
neglect the fermion backreaction on the Q-ball field configuration.
To allow us to neglect  the  fermion  backreaction,  the  bosonic  part  of the
non-scaled Lagrangian (\ref{II:4}) should  be much larger than the Yukawa  term
$G\!\boldsymbol{\phi\cdot}\!\bar{\psi}\boldsymbol{\tau}_{\perp}\psi$.
To estimate these two values, we note  that  the  magnitude of the scalar field
of the Q-ball is $\propto m\Omega g^{-1/2}$, where $\Omega=\left(1 - \omega^{2}
m^{-2} \right)^{1/2}$.
Note that  we return to the  dimensional  quantities  in this and the following
paragraph.
We assume that the dimensionless combination $m^{2}hg^{-2}$ is on the order of
unity.
It then follows that for fixed $\Omega$,  the  contribution of the bosonic part
of the  Lagrangian  (\ref{II:4}) is  $\propto m^{2}g^{-1}$, whereas that of the
Yukawa term is $\propto m g^{-1/2}$.
Hence, the ratio of these  contributions  is $\propto m g^{-1/2}$, which may be
much greater than unity for sufficiently  small $g$. 
For  small $\Omega$, the  bosonic part $\mathcal{L}_{\text{b}}\approx 2^{-1}m^{
4}\Omega^{2} g^{-1}$, whereas the Yukawa term $\mathcal{L}_{\text{Yk}}\approx m
G \Omega L^{-1} g^{-1/2}$, where  $L$  is  the normalised length of the fermion
field.
It follows that the background field  approximation is invalid in the region of
phase frequencies $m \left(1-2gm^{-6}G^{2}L^{-2}\right)\lesssim\left\vert\omega
\right\vert < m$.
This region can be made arbitrarily small if $g \ll 2^{-1} m^{6} G^{-2} L^{2}$;
in  this  case,  the  region  of  non-applicability  of  the  background  field
approximation    corresponds     to     the    so-called    thick-wall   regime
\cite{kus_plb_1997_404, mv, paccetti} of the one-dimensional Q-ball.

We can see that for sufficiently small $g$, the  background field approximation
will be valid over the entire  phase frequency range $\left \vert \omega \right
\vert \in \left(\omega_{\text{tn}}, m\right)$, except for a small region in the
neighbourhood of $\left\vert \omega \right\vert = m$.
From the viewpoint of QFT, however,  we  are  talking about the scattering of a
massless  fermion  of  energy  $\varepsilon$  on  a  quantised  Q-ball  of mass
$M_{\text{Qb}}$.
For the background field approximation to be valid, we  must neglect the recoil
of the Q-ball in this scattering.
It follows that the fermion energy $\varepsilon$ should  be  much less than the
Q-ball mass $M_{\text{Qb}}$.
In the leading order in $g$, the Q-ball mass $M_{\text{Qb}}\propto m^{3}g^{-1}$,
and  hence the condition  $\varepsilon \ll M_{\text{Qb}}$  can be satisfied for
sufficiently small $gm^{-2}$ and fixed $\Omega$.

Although the Q-ball solution in Eq.~(\ref{II:14}) is time-dependent, it remains
invariant under the combined action of the time translation and $SO(2)$ rotation
about the third isotopic axis:
\begin{equation}
\exp \left( -i\delta \omega T_{3}\right) \boldsymbol{\phi }_{\text{Qb}}
\left(t + \delta, x\right) =
\boldsymbol{\phi}_{\text{Qb}}\left(t, x\right),                   \label{III:1}
\end{equation}
where the generator $T_{3} = - i \epsilon_{3 a b}$.
It follows that $\boldsymbol{\phi}_{\text{Qb}}\left(t, x\right)$ vanishes under
the action of the  operator $\partial_{t} - i\omega T_{3}$ in the same way as a
time-independent field  vanishes  under  the  action  of  the  time  derivative
$\partial_{t}$.
Hence, in the background field  of  the  Q-ball,  fermionic  wave functions are
eigenfunctions  of  the  operator $D_{t} = \partial_{t} - i\omega t_{3}$, where
$t_{3} = \tau_{3}/2$.
Using the equation $i D_{t}\psi=\varepsilon\psi$, we obtain the time dependence
of the components $\psi_{i a}$ of the fermionic wave function
\begin{equation}
\psi =
\begin{pmatrix}
e^{-i \varepsilon_{-} t}\psi _{11}\left( x\right)  & \,
e^{-i \varepsilon_{+} t}\psi _{12}\left( x\right)  \\
e^{-i \varepsilon_{-} t}\psi _{21}\left( x\right)  & \,
e^{-i \varepsilon_{+} t}\psi _{22}\left( x\right)
\end{pmatrix},                                                    \label{III:2}
\end{equation}
where $\varepsilon_{\pm} = \varepsilon \pm \omega /2$.
Substituting Eq.~(\ref{III:2}) into Eq.~(\ref{II:10b}), we  find that the Dirac
equation splits into the two independent subsystems
\begin{equation}
i%
\begin{pmatrix}
\psi _{11}^{\prime } \\
\psi _{22}^{\prime }%
\end{pmatrix}%
=%
\begin{pmatrix}
-\varepsilon_{-} &\, F\left( x\right)  \\
-F\left( x\right)  &\, \varepsilon_{+}
\end{pmatrix}%
\begin{pmatrix}
\psi _{11} \\
\psi _{22}%
\end{pmatrix}                                                     \label{III:3}
\end{equation}
and
\begin{equation}
i%
\begin{pmatrix}
\psi _{12}^{\prime } \\
\psi _{21}^{\prime }%
\end{pmatrix}%
=%
\begin{pmatrix}
-\varepsilon_{+} &\, F\left( x\right)  \\
-F\left( x\right)  &\, \varepsilon_{-}
\end{pmatrix}%
\begin{pmatrix}
\psi _{12} \\
\psi _{21}%
\end{pmatrix}.                                                    \label{III:4}
\end{equation}
In Eqs.~(\ref{III:3}) and (\ref{III:4}), the function
\begin{equation}
F\left( x\right) = \frac{2^{3/2}G\Omega }{\sqrt{1+\kappa \cosh \left( 2\Omega
x\right) }},                                                      \label{III:5}
\end{equation}
where the parameter $\kappa=\left(1-\left(16/3\right)h\Omega^{2}\right)^{1/2}$.

The system (\ref{III:3}) contains only the diagonal components  $\psi_{11}$ and
$\psi_{22}$, whereas  the  system  (\ref{III:4}) contains only the antidiagonal
components $\psi_{12}$ and $\psi_{21}$.
To explain this, we introduce  the  operator  $T = \gamma_{5} \otimes \tau_{3}$
and denote the  diagonal  and  antidiagonal parts of the matrix $\psi_{i a}$ as
$\psi_{\text{d}}$ and $\psi_{\text{a}}$, respectively:
\begin{equation}
\psi_{\text{d}} =
\begin{pmatrix}
\psi _{11} & 0 \\
0 & \psi _{22}
\end{pmatrix}
,\quad \psi_{\text{a}} =
\begin{pmatrix}
0 & \psi _{12} \\
\psi _{21} & 0
\end{pmatrix}.                                                    \label{III:6}
\end{equation}
It is readily seen that
\begin{equation}
T\psi_{\text{d}} = \psi_{\text{d}}, \quad
T\psi_{\text{a}} = -\psi_{\text{a}},                              \label{III:7}
\end{equation}
and hence $\psi_{\text{d}}$  and $\psi_{\text{a}}$ are the eigenmatrices of the
operator $T$.
At the same time, it can be shown that the operator $T$ commutes with the Dirac
Hamiltonian
\begin{equation}
H_{D}=\alpha \mathbb{\otimes I}\left( -i\partial _{x}\right) +
G\beta \mathbf{\otimes }\boldsymbol{\phi}_{\text{Qb}}\!
\boldsymbol{\cdot}\!\boldsymbol{\tau}_{\perp},                    \label{III:8}
\end{equation}
where $\alpha = \gamma^{0} \gamma^{1} = \sigma_{3}$  and  $\beta = \gamma^{0} =
\sigma_{1}$.
It follows from this and Eq.~(\ref{III:7})  that the Hamiltonian $H_{D}$ cannot
mix $\psi_{\text{d}}$ and  $\psi_{\text{a}}$, and this results in the splitting
of the Dirac equation into the  two  independent  subsystems  (\ref{III:3}) and
(\ref{III:4}).

Eq.~(\ref{II:13b}) tells us that the parity  transformation switches a diagonal
fermionic state into an antidiagonal one:
\begin{equation}
\psi _{\text{d}}\left( t,x\right) \overset{P}{\longrightarrow }\psi_{\text{a
}}^{P}\left( t,x\right) = \eta_{P}\gamma^{0}\otimes \mathbb{I}\psi_{\text{d
}}\left( t,-x\right).                                             \label{III:9}
\end{equation}
It follows from Eq.~(\ref{III:9}) that for  given  values  of $\varepsilon$ and
$\omega$, the diagonal and  antidiagonal fermionic states are connected to each
other via the unitary $P$-transformation.
The  unitary  $P$-transformation  keeps   the   magnitudes   of  the  incident,
transmitted, and reflected fermionic fluxes unchanged.
Hence, the reflection and  transmission coefficients do not change when passing
from the diagonal to antidiagonal states.
Next, let us denote the transformation  in Eq.~(\ref{II:14}) by the symbol $R$.
The $R$ transformation changes  the  sign  of  the phase frequency $\omega$ and
transforms  a  diagonal  (antidiagonal)  fermionic  state  to  an  antidiagonal
(diagonal) one:
\begin{equation}
\psi _{\text{d,a}}\left( t,x,-\omega \right) \overset{R}{\longrightarrow}
\psi _{\text{a,d}}^{R}\left( t,x,\omega \right) =\mathbb{I}\otimes \tau
_{1}\psi _{\text{d,a}}\left( t,x,-\omega \right).                \label{III:10}
\end{equation}
We see that the  diagonal  (antidiagonal) fermionic states corresponding to the
phase  frequency  $-\omega$   are  unitarily  equivalent  to  the  antidiagonal
(diagonal) fermionic states  corresponding  to  the  phase  frequency $\omega$.
As in  the  previous  case,  unitary  transformation  (\ref{III:10}) keeps  the
magnitudes  of  the  incident,  transmitted,  and  reflected  fermionic  fluxes
unchanged, and hence cannot change the reflection and transmission coefficients.
It follows from the above that  in  the  study of fermion-Q-ball scattering, it
is sufficient to limit ourselves to the case of  positive $\omega$ and diagonal
fermionic states.

The system (\ref{III:3}) describes  the  scattering  of  the diagonal fermionic
states $\psi_{\text{d}}$.
It can be shown that it is equivalent to the second-order differential equation
\begin{flalign}
&\psi _{11}^{\prime \prime }(x)+\left( i\omega +\dfrac{\kappa \Omega \sinh
\left( 2\Omega x\right) }{1+\kappa \cosh \left( 2\Omega x\right) }\right)
\psi _{11}^{\prime }(x) +
\biggl( \varepsilon _{-}\varepsilon _{+}-\dfrac{\Omega \left( 8G^{2}\Omega
+i\varepsilon _{-}\kappa \sinh \left( 2\Omega x\right) \right) }{1+\kappa
\cosh \left( 2\Omega x\right) }\biggr) \psi _{11}(x) = 0         \label{III:11}
\end{flalign}
together with the differential relation
\begin{equation}
\psi _{22}\left( x\right) = F\left(x\right)^{-1}
\left(i\psi_{11}^{\prime}(x)+\varepsilon_{-}\psi_{11}(x)\right). \label{III:12}
\end{equation}

The structure  of   Eq.~(\ref{III:11})  becomes  clearer  if  we  eliminate the
hyperbolic functions.
In order to do this, we change to a new independent variable
\begin{equation}
\xi = \frac{1}{2}\frac{\left( \sqrt{1-\kappa }+\sqrt{1+\kappa }\right)\left(
1-\tanh \left( \Omega x\right) \right) }{\sqrt{1+\kappa }-\sqrt{1-\kappa }
\tanh \left( \Omega x\right)}.                                   \label{III:13}
\end{equation}
Written in terms of this new variable $\xi$,  Eq.~(\ref{III:11}) takes the form
\begin{flalign}
&\psi _{11}^{\prime \prime }(\xi )+\frac{1}{2}\left( \frac{1}{\xi -a}+\frac{
\Omega -i\omega }{\Omega \xi }+\frac{\Omega +i\omega }{\Omega (\xi -1)}
\right) \psi _{11}^{\prime }(\xi)                                \nonumber
 \\
&+\frac{1}{4\Omega (\xi -1)\xi }\left( \frac{8(1-2a)G^{2}\Omega -i\varepsilon
_{-}}{\xi -a} -\frac{\varepsilon _{-}(\varepsilon _{+}-i\Omega )}{\Omega \xi }+
\frac{\varepsilon _{-}(\varepsilon _{+}+i\Omega )}{\Omega (\xi -1)}\right)
\psi _{11}(\xi ) = 0,                                            \label{III:14}
\end{flalign}
where the parameter
\begin{equation}
a=\frac{1}{2}\left(1+\left( 1-\kappa ^{2}\right)^{-1/2}\right) = \frac{1}{2}
+\frac{\sqrt{3}}{8\sqrt{h}\Omega }.                              \label{III:15}
\end{equation}
Eq.~(\ref{III:14}) has four regular singularities  located at the points $\xi =
0$, $\xi = 1$, $\xi = a$, and $\xi = \infty$.
It  follows  that  in  the  neighbourhoods  of  these  points,  the solution to
Eq.~(\ref{III:14}) can  be  expressed  in  terms  of  the  local Heun functions
\cite{Ronveaux, DLMF}.

Using Eqs.~(\ref{III:11})  and  (\ref{III:12}), it can easily be shown that the
component $\psi_{11}(t,x)\propto \exp\left(-i \varepsilon_{-}\left(t - x\right)
\right)$ as $x \rightarrow \pm \infty$.
At the same time, the component $\psi_{22}(t,x)\propto\exp\left(-i\varepsilon_{
+}\left(t + x\right)\right)$ as $x \rightarrow \pm \infty$.
It follows  that far from the Q-ball, the component $\psi_{11}$  corresponds to
a right-chiral massless fermion  moving  to  the  right,  whereas the component
$\psi_{22}$ corresponds  to  a left-chiral massless fermion moving to the left.
Let the right-moving fermionic  wave  $\psi_{11}$  fall  on the Q-ball from the
left.
The  fermion-Q-ball  interaction   results   in  the  transmitted  right-moving
fermionic  wave  $\psi_{11}$   as  $x \rightarrow +\infty$,  and  the reflected
left-moving fermionic wave $\psi_{22}$ as  $x \rightarrow -\infty$.

The solution to Eq.~(\ref{III:14}) corresponding  to  the  transmitted fermionic
wave is
\begin{equation}
\psi_{11}(\xi) = (1-\xi)^{\frac{i \varepsilon_{-}}{2\Omega}}
\left(a-\xi \right)^{\frac{1}{2}}
\xi^{-\frac{i\varepsilon_{-} }{2\Omega }}
 Hl\left[ a,q_{\text{tr}},\alpha_{\text{tr}},\beta_{\text{tr}
},\gamma_{\text{tr}},\delta_{\text{tr}},\xi \right],             \label{III:16}
\end{equation}
where the parameters
\begin{subequations}                                             \label{III:17}
\begin{eqnarray}
\alpha _{\text{tr}} &=&1, \\
\beta _{\text{tr}} &=&\frac{1}{2}, \\
\gamma _{\text{tr}} &=&\frac{1}{2}-\frac{i}{2}\frac{\varepsilon
_{+}+\varepsilon _{-}}{\Omega }, \\
\delta _{\text{tr}} &=&\frac{1}{2}+\frac{i}{2}\frac{\varepsilon
_{+}+\varepsilon _{-}}{\Omega }, \\
q_{\text{tr}} &=&\frac{1}{4}+2(2a-1)G^{2}-\frac{i}{4}\frac{
\varepsilon_{+} +\varepsilon_{-} }{\Omega},
\end{eqnarray}
\end{subequations}
and we use the notation $Hl\left(a, q; \alpha, \beta,\gamma,\delta ;\xi\right)$
for the six-parameter local Heun function \cite{Ronveaux, DLMF}.
Turning to the variable $x$ and using the properties of the local Heun function
\cite{Ronveaux, DLMF}, we obtain  the  leading  term  of the asymptotics of the
transmitted fermionic wave as
\begin{equation}
\psi _{11}\left( x\right) \sim a^{\frac{1}{2}}e^{\frac{i\varepsilon _{-}}{
4\Omega}\ln \left(1-\frac{1}{a}\right)}e^{i\varepsilon_{-}x}     \label{III:18}
\end{equation}
as $x \rightarrow +\infty$.

The local Heun function $Hl\left(a, q; \alpha, \beta, \gamma,\delta;\xi\right)$
is analytic, and is equal to one at the regular singular point $\xi = 0$.
Hence, it can be expanded in a Taylor series about this point.
In the complex $\xi$-plane, the  radius  of convergence of this series is equal
to $\min \left(1, a\right) = 1$,  as  it  follows  from Eq.~(\ref{III:15}) that
$a > 1$.
In this case, the local Heun function $Hl\left(a, q;\alpha,\beta,\gamma,\delta;
\xi \right)$ can  be  analytically continued from the  unit disk to  the  whole
complex plane  with  the  branch  cut  discontinuity $\left[1, \infty \right)$,
and its values are indeterminate at the regular singular points  $\xi = 1$  and
$\xi = a$.
It also follows   from   Eq.~(\ref{III:13})  that  $\xi  \rightarrow  1$  as $x
\rightarrow -\infty$.
Hence, we cannot use  Eq.~(\ref{III:16})  to  describe  the  incident fermionic
wave far to the left of the Q-ball.

To describe the incident fermionic wave, we  must use the local Heun functions,
which are well-defined at the point $\xi = 1$.
Knowing  the   prefactor   (the   product   of   the   first  three  factors in
Eq.~(\ref{III:16})) and  parameters (\ref{III:17}),  and   using   the symmetry
properties of Heun's  equation \cite{Ronveaux, DLMF}, we can  write the general
form of  the  local  solution  to  Eq.~(\ref{III:14})  in  the neighbourhood of
$\xi = 1$ as
\begin{eqnarray}
\psi _{11}(\xi ) &=&
c_{1}(1-\xi )^{\frac{i\varepsilon_{-}}{2\Omega }}\left(a-\xi\right)^{\frac{1
}{2}}\xi^{-\frac{i\varepsilon_{-}}{2\Omega }}
Hl\left[1-a,q_{\text{in}},\alpha_{\text{in}},\beta_{\text{in}
},\gamma _{\text{in}},\delta _{\text{in}},1-\xi \right]          \nonumber
 \\
&&+c_{2}(1-\xi )^{\frac{1}{2}-\frac{i\varepsilon _{+}}{2\Omega }}
\left( a-\xi \right) ^{\frac{1}{2}}
\xi ^{-\frac{i\varepsilon _{-}}{2\Omega }}
Hl\left[ 1-a,q_{\text{rf}},\alpha _{\text{rf}},\beta _{\text{rf}
},\gamma _{\text{rf}},\delta _{\text{rf}},1-\xi \right],         \label{III:19}
\end{eqnarray}
where the parameters
\begin{subequations}                                             \label{III:20}
\begin{eqnarray}
\alpha _{\text{in}} &=&1, \\
\beta _{\text{in}} &=&\frac{1}{2}, \\
\gamma _{\text{in}} &=&\frac{1}{2}+\frac{i}{2}\frac{\varepsilon
_{+}+\varepsilon _{-}}{\Omega }, \\
\delta _{\text{in}} &=&\frac{1}{2}-\frac{i}{2}\frac{\varepsilon
_{+}+\varepsilon _{-}}{\Omega }, \\
q_{\text{in}} &=&\frac{1}{2}-q_{\text{tr}},
\end{eqnarray}
\end{subequations}
and
\begin{subequations}                                             \label{III:21}
\begin{eqnarray}
\alpha _{\text{rf}} &=&\frac{3}{2}-\frac{i}{2}\frac{\varepsilon
_{+}+\varepsilon _{-}}{\Omega }, \\
\beta _{\text{rf}} &=&1-\frac{i}{2}\frac{\varepsilon _{+}+\varepsilon _{-}}{%
\Omega }, \\
\gamma _{\text{rf}} &=&\frac{3}{2}-\frac{i}{2}\frac{\varepsilon
_{+}+\varepsilon _{-}}{\Omega }, \\
\delta _{\text{rf}} &=&\frac{1}{2}-\frac{i}{2}\frac{\varepsilon
_{+}+\varepsilon _{-}}{\Omega }, \\
q_{\text{rf}} &=&\frac{6-a}{4}+\frac{\left( a-1\right) \left( \varepsilon
_{+}+\varepsilon _{-}\right) ^{2}}{4\Omega ^{2}}                 \nonumber
 \\
&&+i\frac{\left( 2a-5\right) \left( \varepsilon _{+}+\varepsilon _{-}\right)
}{4\Omega }-q_{\text{tr}}.
\end{eqnarray}
\end{subequations}

The solutions in Eqs.~(\ref{III:16}) and (\ref{III:19}) have a common domain of
analyticity.
To determine the  coefficients  $c_{1}$  and  $c_{2}$ in Eq.~(\ref{III:19}), we
need  to  equate  Eqs.~(\ref{III:16})  and  (\ref{III:19})  as  well  as  their
derivatives in $\xi$ at any point in their common  domain  of  analyticity; the
values of the coefficients  will  not  depend  on  the  specific choice of this
point.
To simplify the formulae, we  choose  the  symmetric  point  $\xi = 1/2$ as the
matching point.
As a result, we  obtain  the following expressions for the coefficients $c_{1}$
and $c_{2}$:
\begin{eqnarray}
c_{1} &=&\frac{\chi_{\text{tr}}\left(1/2\right) }{\chi_{\text{in}}\left(
1/2\right)}-\Omega \frac{\chi_{\text{rf}}\left(1/2\right) W_{\text{ti}
}\left(1/2\right)}{\chi_{\text{in}}\left( 1/2\right) V\left( 1/2\right)},
                                                                \label{III:22a}
\\
c_{2} &=&2^{\frac{1}{2}-\frac{i}{2}\frac{\varepsilon _{+}+\varepsilon _{-}}{
\Omega }}\Omega \frac{W_{\text{ti}}\left( 1/2\right) }{V\left( 1/2\right) },
                                                                \label{III:22b}
\end{eqnarray}
where the functions
\begin{subequations}                                             \label{III:23}
\begin{eqnarray}
\chi _{\text{in}}\left( \xi \right)  &=&Hl\left[ 1-a,q_{\text{in}},\alpha _{
\text{in}},\beta _{\text{in}},\gamma _{\text{in}},\delta _{\text{in}},1-\xi
\right] , \\
\chi _{\text{rf}}\left( \xi \right)  &=&Hl\left[ 1-a,q_{\text{rf}},\alpha _{
\text{rf}},\beta _{\text{rf}},\gamma _{\text{rf}},\delta _{\text{rf}},1-\xi
\right] , \\
\chi _{\text{tr}}\left( \xi \right)  &=&Hl\left[ a,q_{\text{tr}},\alpha _{
\text{tr}},\beta _{\text{tr}},\gamma _{\text{tr}},\delta _{\text{tr}},\xi
\right],
\end{eqnarray}
\end{subequations}
the Wronskians
\begin{subequations}                                             \label{III:24}
\begin{eqnarray}
W_{\text{ri}}\left( \xi \right)  &=&\chi _{\text{rf}}\left( \xi \right) \chi
_{\text{in}}^{\prime }\left( \xi \right) -\chi _{\text{rf}}^{\prime }\left(
\xi \right) \chi _{\text{in}}\left( \xi \right),  \\
W_{\text{ti}}\left( \xi \right)  &=&\chi _{\text{tr}}\left( \xi \right) \chi
_{\text{in}}^{\prime }\left( \xi \right) -\chi _{\text{tr}}^{\prime }\left(
\xi \right) \chi _{\text{in}}\left( \xi \right),
\end{eqnarray}
\end{subequations}
and the combination
\begin{equation}
V\left( \xi \right) =\Omega W_{\text{ri}}\left( \xi \right) +\left( \Omega
-i\left( \varepsilon _{+}+\varepsilon _{-}\right) \right) \chi _{\text{in}
}\left( \xi \right) \chi _{\text{rf}}\left(\xi \right).          \label{III:25}
\end{equation}

As $x  \rightarrow  -\infty$, the  asymptotics  of  $\psi_{11}$  takes the form
\begin{equation}
\psi _{11} \sim c_{1}\left( a-1\right) ^{\frac{1}{2}}e^{\frac{i\varepsilon
_{-}}{4\Omega}\ln\left(1-\frac{1}{a}\right)}e^{i\varepsilon_{-}x}
 +c_{2}\left( a-1\right) ^{\frac{3}{4}}a^{-\frac{1}{4}}e^{-\frac{
i\varepsilon _{+}}{4\Omega }\ln \left( 1-\frac{1}{a}\right)}e^{\Omega
x}e^{-i\varepsilon_{+}x}.                                        \label{III:26}
\end{equation}
We see that  as  $x   \rightarrow   -\infty$,  $\psi_{11}$  is  the  sum of the
right-moving wave that is $\propto  c_{1}$   (the  incident fermionic wave) and
the exponentially damped wave that is $\propto c_{2}$.
Using Eqs.~(\ref{III:12}) and (\ref{III:26}),  we  can ascertain the asymptotic
behaviour of the $\psi_{22}$  component  as $x \rightarrow -\infty$ as follows:
\begin{equation}
\psi_{22} \sim \frac{c_{2}}{2^{3/2}G}\frac{a-1}{\left(2a-1\right)^{1/2}}
\frac{\varepsilon _{+}+\varepsilon _{-}+i\Omega }{\Omega}
e^{-\frac{i \varepsilon _{+}}{4\Omega }\ln \left( 1-a^{-1}\right)
}e^{-i\varepsilon_{+}x}.                                         \label{III:27}
\end{equation}
We see that  as  $x \rightarrow -\infty$, $\psi_{22}$  is  the left-moving wave
(the reflected fermionic wave) which is $\propto  c_{2}$.

In the background field of the Q-ball, fermions  become asymptotically free and
massless as $x \rightarrow \pm \infty$.
The free massless fermions can be characterised  by their energy $\varepsilon_{
\pm}$, isospin projection $I_{3}$, and chirality $\pm 1$.
Using Dirac's notations, we  denote  the  asymptotic  states  of  the incident,
transmitted  and  reflected   fermions  as  $\left\vert \varepsilon_{-}, 1/2, R
\right\rangle^{\text{(in)}}$, $\left\vert\varepsilon_{-},1/2, R \right\rangle^{
\text{(out)}}$, and $\left\vert\varepsilon_{+},-1/2,L\right\rangle^{\text{(out)
}}$, respectively, where $R$  ($L$)  denotes  positive (negative) chirality.
Following  the   standard   method   of   scattering   theory  \cite{LandauIII,
 Goldberger, Taylor},  we   shall   use   the  $S$  operator  to  describe  the
fermion-Q-ball scattering.
Acting on  the  state $\left\vert \varepsilon_{-}, 1/2, R \right\rangle^{\text{
(in)}}$, the $S$ operator  turns  it  into  a  linear combination of the states
$\left\vert\varepsilon_{-},1/2, R \right\rangle^{\text{(out)}}$ and $\left\vert \varepsilon_{+},-1/2,L\right\rangle^{\text{(out)}}$:
\begin{equation}
S\left\vert \varepsilon_{-},1/2,R\right\rangle^{\text{(in)}}
= S_{1/2,1/2}\left\vert\varepsilon_{-},1/2,R\right\rangle^{\text{(out)}}
+ S_{-1/2,1/2}\left\vert \varepsilon_{+},-1/2,L\right\rangle^{\text{(out)}},
                                                                 \label{III:28}
\end{equation}
where the $S$-matrix elements
\begin{subequations}                                             \label{III:29}
\begin{eqnarray}
S_{1/2,1/2}&=&^{\text{(out)}}\!\left\langle \varepsilon_{-},1/2,R\left\vert S
\right\vert \varepsilon_{-},1/2,R\right\rangle^{\text{(in)}},   \label{III:29a}
\\
S_{-1/2,1/2}&=&^{\text{(out)}}\!\left\langle\varepsilon_{+},-1/2,L\left\vert S
\right\vert \varepsilon _{-},1/2,R\right\rangle^{\text{(in)}}.  \label{III:29b}
\end{eqnarray}
\end{subequations}

Using asymptotic  forms  (\ref{III:18}), (\ref{III:26}), and (\ref{III:27}), we
can  write  the  $S$-matrix  elements  included  in  Eq.~(\ref{III:28}) as
\begin{equation}
S_{1/2,1/2}=\left( 1-a^{-1}\right) ^{-1/2}c_{1}^{-1}             \label{III:30}
\end{equation}
and
\begin{equation}
S_{-1/2,1/2} = 2^{-3/2}c_{2}c_{1}^{-1}G^{-1}\left( \frac{a-1}{2a-1}\right)
^{1/2}\left( 2\varepsilon \Omega ^{-1}+i\right) e^{-i\frac{\varepsilon}
{2\Omega }\ln \left( 1-a^{-1}\right)}.                           \label{III:31}
\end{equation}
The $S$-matrix elements (\ref{III:30}) and (\ref{III:31}) satisfy the unitarity
condition
\begin{equation}
\left\vert S_{-1/2,1/2}\right\vert ^{2}+\left\vert S_{1/2,1/2}\right\vert
^{2} = 1.                                                        \label{III:32}
\end{equation}

Substituting asymptotic forms (\ref{III:18}), (\ref{III:26}), and (\ref{III:27})
into the expression for the spatial component  of  the fermion current given in
Eq.~(\ref{II:9}), we can calculate the transmission and reflection coefficients:
\begin{equation}
T=\frac{j_{\text{tr}}}{j_{\text{in}}}=\left\vert c_{1}\right\vert ^{-2}\frac{
a}{a-1}                                                          \label{III:33}
\end{equation}
and
\begin{equation}
R=\frac{\left\vert j_{\text{rf}}\right\vert }{j_{\text{in}}}=\frac{
\left\vert c_{2}\right\vert ^{2}}{\left\vert c_{1}\right\vert ^{2}}\frac{1}
{8G^{2}}\frac{a-1}{2a-1}\left( 1+4\varepsilon ^{2}\Omega ^{-2}\right).
                                                                 \label{III:34}
\end{equation}
It  follows  from   Eqs.~(\ref{III:30}),  (\ref{III:31}),  (\ref{III:33}),  and
(\ref{III:34}) that  the   transmission  and  reflection  coefficients  are the
squared magnitudes of the corresponding $S$-matrix elements:
\begin{equation}
T = \left\vert S_{1/2,1/2}\right\vert ^{2}\;\;\text{and}\;\; R = \left\vert
S_{-1/2,1/2}\right\vert^{2}.                                     \label{III:35}
\end{equation}
Eqs.~(\ref{III:32}) and (\ref{III:35}) then tell  us  that  for given values of
$\omega$  and  $\varepsilon$  (recall that $\varepsilon = \varepsilon_{\pm }\mp
\omega /2$), the transmission and reflection coefficients satisfy the unitarity
condition
\begin{equation}
T\left(\varepsilon,\omega\right)+R\left(\varepsilon,\omega \right) = 1.
                                                                 \label{III:36}
\end{equation}

In the process of scattering, the transmitted fermionic  wave acquires a  phase
shift  $\delta$  with  respect  to  the  incident fermionic wave.
From Eqs.~(\ref{III:18}), (\ref{III:26}),  and  (\ref{III:30}) it  follows that
\begin{equation}
\delta \left( \varepsilon ,\omega \right) =-\arg \left[ c_{1}\left(
\varepsilon ,\omega \right) \right] =\arg \left[ S_{1/2,1/2}\left(
\varepsilon ,\omega \right) \right].                             \label{III:37}
\end{equation}

Eqs.~(\ref{III:26})  and  (\ref{III:27})  tell  us  that  there  are  no  bound
diagonal fermionic  states  $\psi_{\text{d}}$  in the  background  field of the
Q-ball.
This is because there are   no  diagonal  fermionic  states  for which both the
$\psi_{11}$ and $\psi_{22}$ components decrease exponentially as $x \rightarrow
\pm \infty$.
A  similar   situation    occurs   for   the   antidiagonal   fermionic  states
$\psi_{\text{a}}$.
In the model under consideration, fermions acquire mass only through the Yukawa
interaction.
As $x\rightarrow \pm\infty$, the Q-ball's field $\boldsymbol{\phi}_{\text{Qb}}$
tends to zero exponentially,  and thus fermions become asymptotically massless.
The absence of a mass gap  in  the  spectrum  of the Dirac Hamiltonian makes it
impossible for fermionic bound states to exist  in the  background field of the
Q-ball.

\section{Evaporation of the Q-ball}                          \label{sec:IV}

It was shown in Ref.~\cite{ccgm}  that  the  Yukawa  interaction  of the scalar
field of a Q-ball with massless fermions  leads  to  evaporation of the Q-ball.
In our case, this means that the energy  and  Noether  charge of the Q-ball are
carried away by a flux of fermion-antifermion pairs.
The evaporation of the  Q-ball  is  possible  only  when  the  energy parameter
$\varepsilon \in \left( -\omega /2,\omega /2\right)$.
In this case, the parameter $\varepsilon _{-} = \varepsilon -\omega /2$ becomes
negative, and  therefore in  Eq.~(\ref{III:2}),  the components $\psi_{11}$ and
$\psi_{21}$ correspond to antifermionic states.

As $\omega \rightarrow\omega_{\text{tn}}$, the Q-ball passes into the thin-wall
regime.
In  this  regime,  the  energy,  Noether   charge,  and  spatial  size  of  the
one-dimensional Q-ball increase indefinitely in the limit of $\omega \rightarrow
\omega_{\text{tn}}$, and  the  energy  and  charge  densities  become spatially
homogeneous within the Q-ball, except for two thin  boundary transition layers.
It was shown in Ref.~\cite{ccgm} that in the thin-wall  regime,  the production
of pairs cannot occur in  the  interior of the  Q-ball,  but  only  in the thin
transition layer at its boundary.
In our case, this  is  because the scalar-fermion Yukawa interaction shifts the
point to which we fill the Dirac sea.
Specifically, for the  components $\psi_{12}$ and $\psi_{22}$, the Dirac sea is
filled to the level $\omega/2$ and is overflowed.
In contrast, for  the  components $\psi_{11}$ and $\psi_{21}$, the Dirac sea is
filled to the level $-\omega/2$ and is underflowed.
As a result, no fermion-antifermion  pairs  with a total energy of $\omega$ can
be produced in the interior of  the  Q-ball, and evaporation becomes impossible
there.

However, evaporation is possible  in  the  thin  transition layer at the Q-ball
boundary.
In  Ref.~\cite{ccgm},  the  evaporation rate was calculated based on a  leading
order semi-classical approximation, in which  massless fermions were considered
in the presence of the classical background field of a Q-ball.
A detailed derivation  of  the evaporation rate for the one-dimensional case is
given in Ref.~\cite{clark}.
In our case, the evaporation  rate  of  the  Noether  charge  of  the Q-ball is
\begin{equation}
\frac{dQ}{dt}=\frac{2}{\pi}m\int\limits_{0}^{\frac{
\tilde{\omega }}{2}}R\bigl( \tilde{\varepsilon},\tilde{\omega},
\tilde{h},\tilde{G}\bigr) d\tilde{\varepsilon},                    \label{IV:1}
\end{equation}
where $R$ is the reflection coefficient and $m$ is the mass of the scalar field
$\boldsymbol{\phi}$.
In the  discussion  in this section, the  dimensionless  parameters (defined by
Eq.~(\ref{II:5a})) are marked with a tilde, so that the dimensionless arguments
of  $R$  are  $\tilde{\varepsilon } =  m^{-1}\varepsilon$,   $\tilde{\omega } =
m^{-1}\omega$, $\tilde{h} = m^{2}g^{-2}h$, and $\tilde{G} = g^{-1/2}G$.
In Eq.~(\ref{IV:1}), the  reflection  coefficient $R$  is  calculated using the
general formulae in Sec.~\ref{sec:III}.
Using the relation $dE/dQ = \omega$,  we  can also calculate the rate of energy
loss of the Q-ball
\begin{equation}
\frac{dE}{dt}  = \frac{dE}{dQ}\frac{dQ}{dt}=\omega \frac{dQ}{dt}
= \frac{2}{\pi }\omega m\int\limits_{0}^{\frac{\tilde{\omega }}{2}
}R\bigl(\tilde{\varepsilon},\tilde{\omega},\tilde{h},
\tilde{G}\bigr) d\tilde{\varepsilon}.                              \label{IV:2}
\end{equation}

Using the inequality  $R  <  1$,  which  follows  from  the unitarity condition
(\ref{III:36}), we  obtain  the  upper  bound  on  the  evaporation rate of the
Noether  charge  of  the Q-ball
\begin{equation}
\left. \frac{dQ}{dt}\right\vert_{\max }  =  \frac{
\omega }{\pi }\approx \frac{\omega _{\text{tn}}}{\pi} =
\frac{m}{\pi }\left[ 1-\left( 3/16\right) \left( m^{2}hg^{-2}\right)
^{-1}\right]^{1/2},                                                \label{IV:3}
\end{equation}
where we use the fact that  $\omega\approx\omega_{\text{tn}}$  in the thin-wall
regime.
It is important to note that this upper  bound  does  not  depend on the Yukawa
coupling constant $G$, and is  determined only by the parameters of the bosonic
sector of model (\ref{II:4}).
The combination of Eqs.~(\ref{IV:2}) and (\ref{IV:3})  gives us the upper bound
on the rate of energy loss of the Q-ball
\begin{eqnarray}
\left. \frac{dE}{dt}\right\vert _{\max }  =  \frac{\omega ^{2}}{\pi }\approx
\frac{\omega _{\text{tn}}^{2}}{\pi}
 = \frac{m^{2}}{\pi }\left[ 1-\left( 3/16\right) \left( m^{2}hg^{-2}\right)
^{-1}\right].                                                      \label{IV:4}
\end{eqnarray}

In  the  thin-wall  regime,  the  energy  and   Noether  charge  densities  are
approximately constant in the interior of the Q-ball.
In this regime, the magnitude of the Noether charge density is
\begin{equation}
j_{0} \approx \frac{3}{2}\frac{m g}{h}\left(1-\frac{3}{16}
\frac{g^{2}}{m^{2}h}\right)^{1/2}.                                 \label{IV:5}
\end{equation}
It follows that in the thin-wall regime, the magnitude of the Noether charge of
the Q-ball is
\begin{equation}
 Q \approx  j_{0} L,                                               \label{IV:6}
\end{equation}
where $L$ is the linear size of the one-dimensional Q-ball.
Using Eqs.~(\ref{IV:1}),  (\ref{IV:5}),  and  (\ref{IV:6}),  we  can  find  the
velocity of recession of $L$ as a result of Q-ball evaporation as follows:
\begin{equation}
\frac{dL}{dt}  =  \frac{1}{j_{0}}\frac{dQ}{dt}=\frac{2}{3}\frac{h}{g}
\left( 1-\frac{3}{16}\frac{g^{2}}{m^{2}h}\right) ^{-1/2}
\frac{2}{\pi }\int\limits_{0}^{\frac{\tilde{\omega }}{2}
}R\bigl( \tilde{\varepsilon },\tilde{\omega },\tilde{h},
\tilde{G}\bigr) d\tilde{\varepsilon}.                              \label{IV:7}
\end{equation}
From Eqs.~(\ref{III:36}) and (\ref{IV:7}) it  follows  that  the upper bound on
$dL/dt$ is
\begin{equation}
\left. \frac{dL}{dt}\right\vert_{\max}\approx \frac{2}{3\pi}\frac{h}{g} =
\frac{2\tilde{h}}{3\pi}\frac{g}{m^{2}}.                            \label{IV:8}
\end{equation}

All of the formulae in this  section  were  derived  in  the leading (one-loop)
order of the semi-classical approximation, which is valid in the limit of small
$\hbar$.
In this paper, however, we use the natural units $\hbar = c = 1$.
It can be shown \cite{lee_pang_1992}  that  in  this  case,  the semi-classical
limit is  the  limit  of  small  $g m^{-2}$, where  the remaining dimensionless
combinations $\tilde{\omega}=\omega/m$, $\tilde{h}=m^{2}g^{-2}h$, and $\tilde{G}
= g^{-1/2}G$ are fixed and are $\lesssim 1$.
This is because after scaling as in Eq.~(\ref{II:5a}), the action $S_{T} = \int
\nolimits_{0}^{\frac{2\pi}{\omega}}\int\nolimits_{-\infty}^{+\infty}\mathcal{L}
dxdt$ over the period $T = 2\pi/\omega$ is scaled as
\begin{equation}
S_{T}\left(\omega,m,g,h,G\right) \rightarrow m^{2}g^{-1}
S_{T}\bigl(\tilde{\omega},1,1,\tilde{h},\tilde{G}\bigr).           \label{IV:9}
\end{equation}
It follows  that  as  $m^{2}g^{-1}  \rightarrow  \infty$,  the  contribution of
arbitrary field configurations  to  the  functional  integrals of  QFT  will be
strongly suppressed, due to fast  oscillations  in the exponential factor $\exp
\left[i m^{2} g^{-1} S_{T}\right]$.
In this case, the main contribution to the functional integrals comes from field
configurations  in  infinitesimal  neighbourhoods  of  the  classical  solutions
(stationary points of action), which is the main  feature  of the semi-classical
regime.

As noted above, the existence of the Q-ball is only possible if $\tilde{h}=m^{2
}g^{-2}h>3/16$.
On the other hand, we assume that  the  dimensionless constant $\tilde{h}$ must
be on the order of or less  than  unity  for  the  model  to be consistent from
the viewpoint of QFT.
Eq.~(\ref{IV:8}) then  tells  us  that  in  the semi-classical  limit  of small
$g m^{-2}$, the velocity $dL/dt$ is much smaller than the speed of light $c=1$.

\section{Numerical results}                                       \label{sec:V}

The Dirac equation (\ref{II:10b}),  when  written  in  terms  of dimensionless
quantities (\ref{II:5a}), depends on the three  dimensionless  parameters $h$,
$G$, and $\omega$.
Our main goal is to ascertain the dependence  of  the fermion-Q-ball scattering
on the Yukawa coupling constant $G$ and  phase frequency $\omega$.
For this reason, we fix  the  nonlinear  coupling  constant $h$ to $0.2$.
At this value of $h$, the magnitude of  the  phase  frequency of the Q-ball can
vary in a comparatively wide range $\left(0.25, 1 \right)$.

\begin{figure}[tbp]\center
\includegraphics[width=0.5\textwidth]{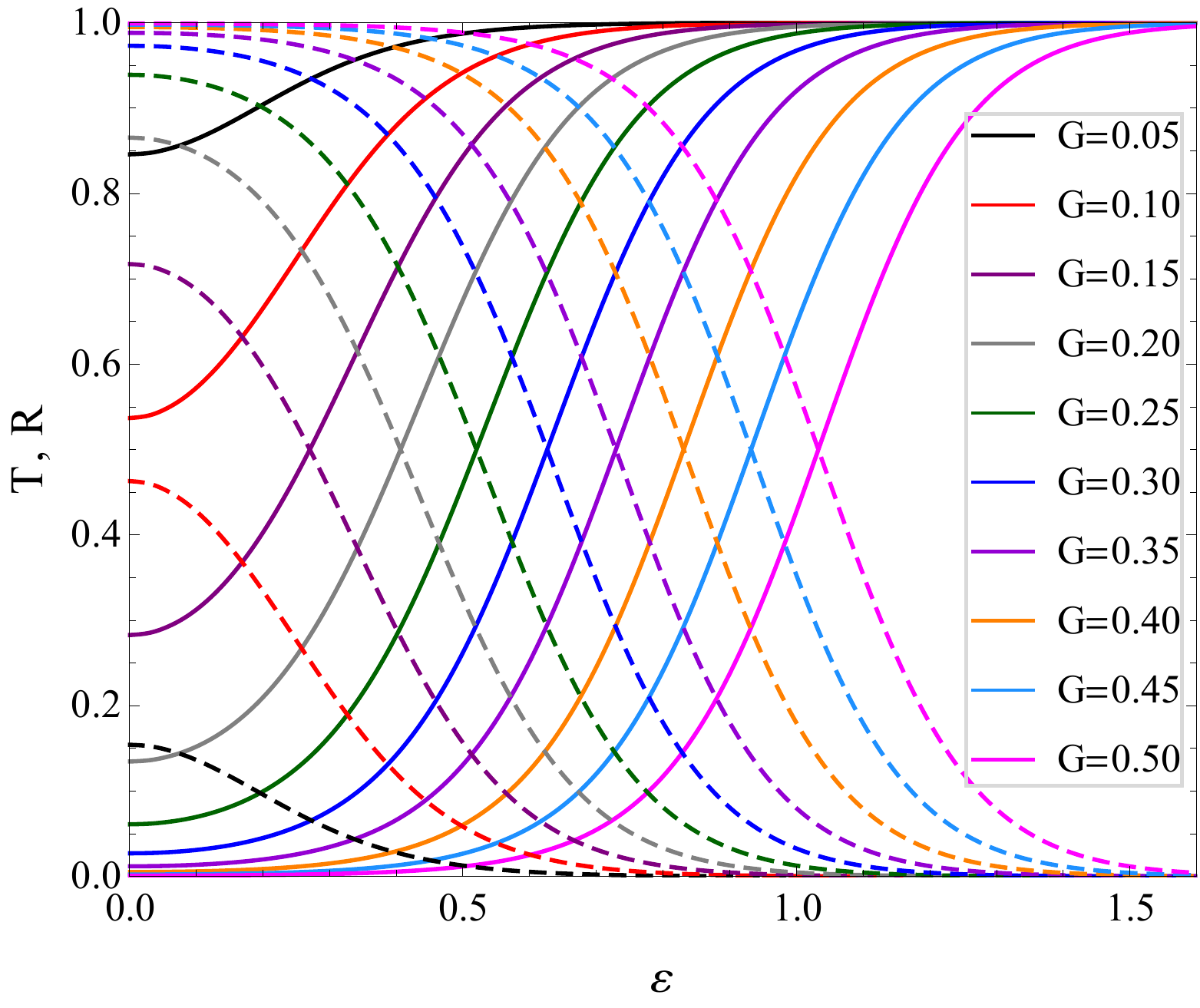}
\caption{Dependence  of  the  transmission  coefficient  $T$ (solid curves) and
reflection coefficient $R$ (dashed curves) on the energy parameter $\varepsilon$
for different values of the Yukawa coupling constant $G$. The curves correspond
to the parameters $h = 0.2$ and $\omega = 0.5$\hspace{1pt}.}
\label{fig:1}
\end{figure}

Figure~\ref{fig:1} illustrates the dependence  of  the transmission coefficient
$T$ and reflection coefficient $R$  on   the  energy  parameter  $\varepsilon =
(\varepsilon_{+}+\varepsilon_{-})/2$ for $h = 0.2$, $\omega=0.5$, and different
values of the Yukawa coupling constant $G$.
It follows from Fig.~\ref{fig:1} that $T\left(\varepsilon \right)$ and $R\left(
\varepsilon \right)$ are even functions of $\varepsilon$.
This symmetry property is a consequence of the invariance of the Dirac equation
(\ref{II:10b}) under the $C$-conjugation (\ref{II:13a}).
In  Fig.~\ref{fig:1},  the  curves  $T\left(\varepsilon \right)$  and  $R\left(
\varepsilon \right)$  corresponding  to  the  same  value  of  $G$  satisfy the
unitarity condition (\ref{III:36}).
Another characteristic property is  that  for  nonzero $G$, the curves $T\left(
\varepsilon \right)$ do not vanish at $\varepsilon = 0$.
Accordingly, the  curves $R\left( \varepsilon \right)$ are different from unity
at $\varepsilon = 0$ and nonzero $G$.
This property is  a  consequence  of  the  non-trivial  time  dependence of the
Q-ball solution in Eq.~(\ref{II:11}),  which  results in the nonconservation of
energy in fermion-Q-ball scattering.
We find that for sufficiently large  $G$,  the  value  of  $T(0)$ tends to zero
exponentially with increasing $G$.
Accordingly, the  value of $R(0)$  tends to unity exponentially in this regime.
As $G \rightarrow 0$, the  value  of  $R(0)$  ($T(0)$)  tends  to  zero (unity)
$\propto G^{2}$.

Let us define  the  parameter  $\varepsilon_{1/2}$  by  the  condition $T\left(
\varepsilon_{1/2} \right) = R\left( \varepsilon_{1/2} \right) = 1/2$.
We also determine the effective Yukawa  mass  of the  fermion in the background
field of the Q-ball as follows:
\begin{equation}
m_{\psi\!,\text{eff}}  =  G\left\vert \phi \left(0\right) \right\vert
= 2^{3/2}G\Omega \left[1 + \left(1 - 16h\Omega^{2}/3\right) ^{1/2}\right]
^{-1/2},                                                            \label{V:1}
\end{equation}
where $\Omega = \left(1 - \omega^{2}\right)^{1/2}$.
We then find   that   in   Fig.~\ref{fig:1}, the  parameter $\varepsilon_{1/2}$
satisfies  the approximate relation
\begin{equation}
\varepsilon_{1/2} \approx m_{\psi ,\text{eff}}                      \label{V:2}
\end{equation}
for the curves $T\left(\varepsilon \right)$  and $R\left( \varepsilon \right)$,
such that $T\left(0\right) \ll 1$ and $1 -R\left(0\right) \ll 1$, respectively.
In particular, the  parameter  $\varepsilon_{1/2}$ is approximately $\propto G$
for these curves.

\begin{figure}[tbp]\center
\includegraphics[width=0.5\textwidth]{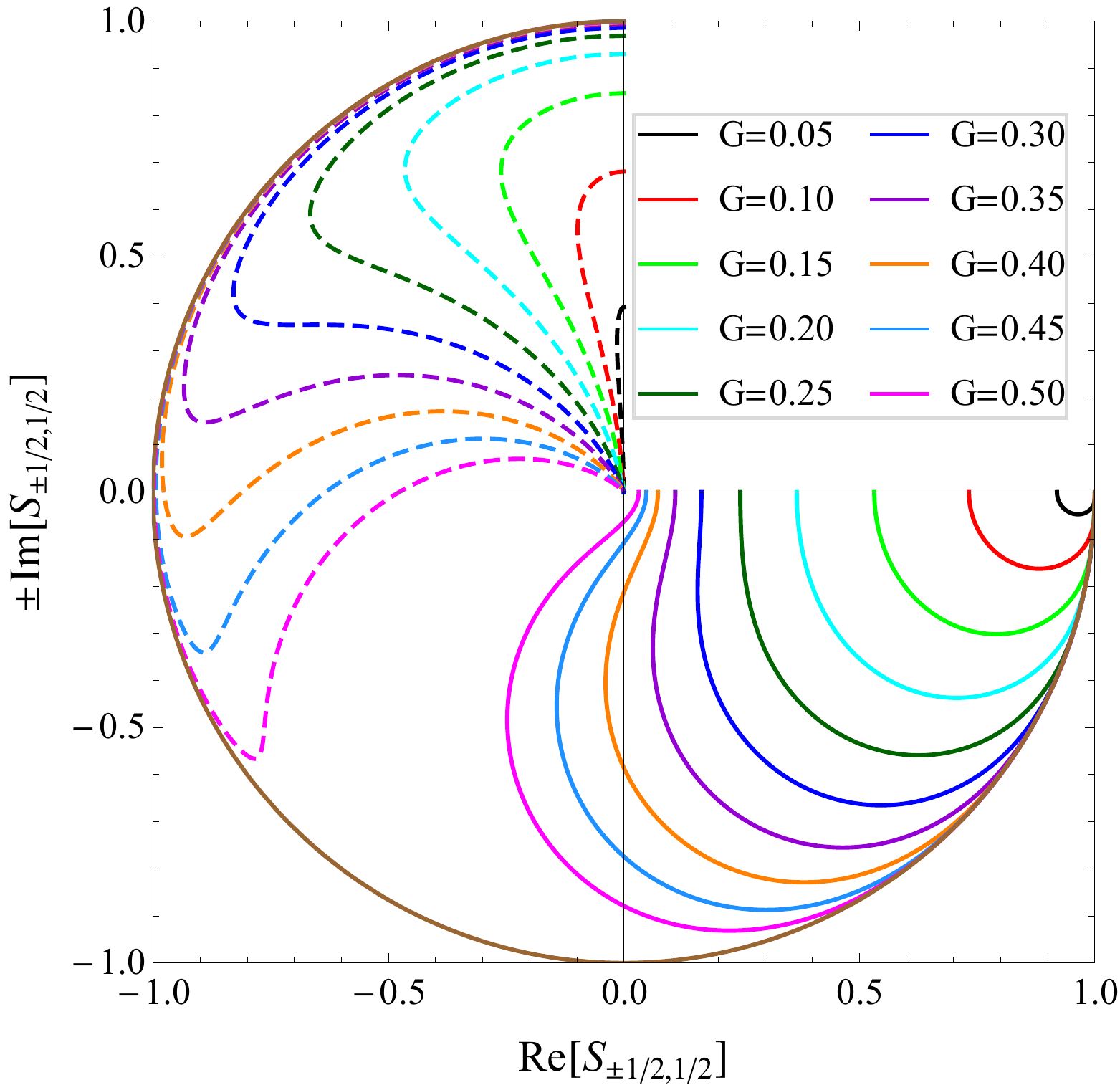}
\caption{Argand diagram  for  the  $S$-matrix  elements  $S_{1/2,\,1/2}$ (solid
curves) and  the complex  conjugate  $S$-matrix  elements  $S^{*}_{-1/2,\,1/2}$
(dashed curves) for different values of  the Yukawa coupling constant $G$.  The
curves correspond to the parameters $h = 0.2$ and $\omega = 0.5$\hspace{1pt}.}
\label{fig:2}
\end{figure}

Figure~\ref{fig:2} presents an  Argand   diagram   for  the  elastic $S$-matrix
elements $S_{1/2,\,1/2}$ and the complex conjugate inelastic $S$-matrix elements
$S^{*}_{-1/2,\,1/2}$ for different values of  the Yukawa coupling constant $G$.
We present the conjugate  $S$-matrix  elements $S^{*}_{-1/2,\,1/2}$ rather than
$S_{-1/2,\,1/2}$ in order to avoid  intersection  between  the solid and dashed
curves.
It  follows  from  Fig.~\ref{fig:2}  that  for  $\varepsilon = 0$,  the elastic
$S$-matrix elements $S_{1/2,\,1/2}$ are real.
They tend  to unity  as $G \rightarrow 0$, and to zero with an increase in $G$.
As $\varepsilon$ increases, the matrix  elements $S_{1/2,\,1/2}$ leave the real
axis and move along the solid curves in Fig.~\ref{fig:2}.
With a further rise in $\varepsilon$, they approach the unitary circle and then
tend to unity as $\varepsilon \rightarrow \infty$.
It follows that when $G$  is  fixed  and  $\varepsilon \rightarrow \infty$, the
fermion-Q-ball interaction becomes negligibly small.

\begin{figure}[tbp]\center
\includegraphics[width=0.5\textwidth]{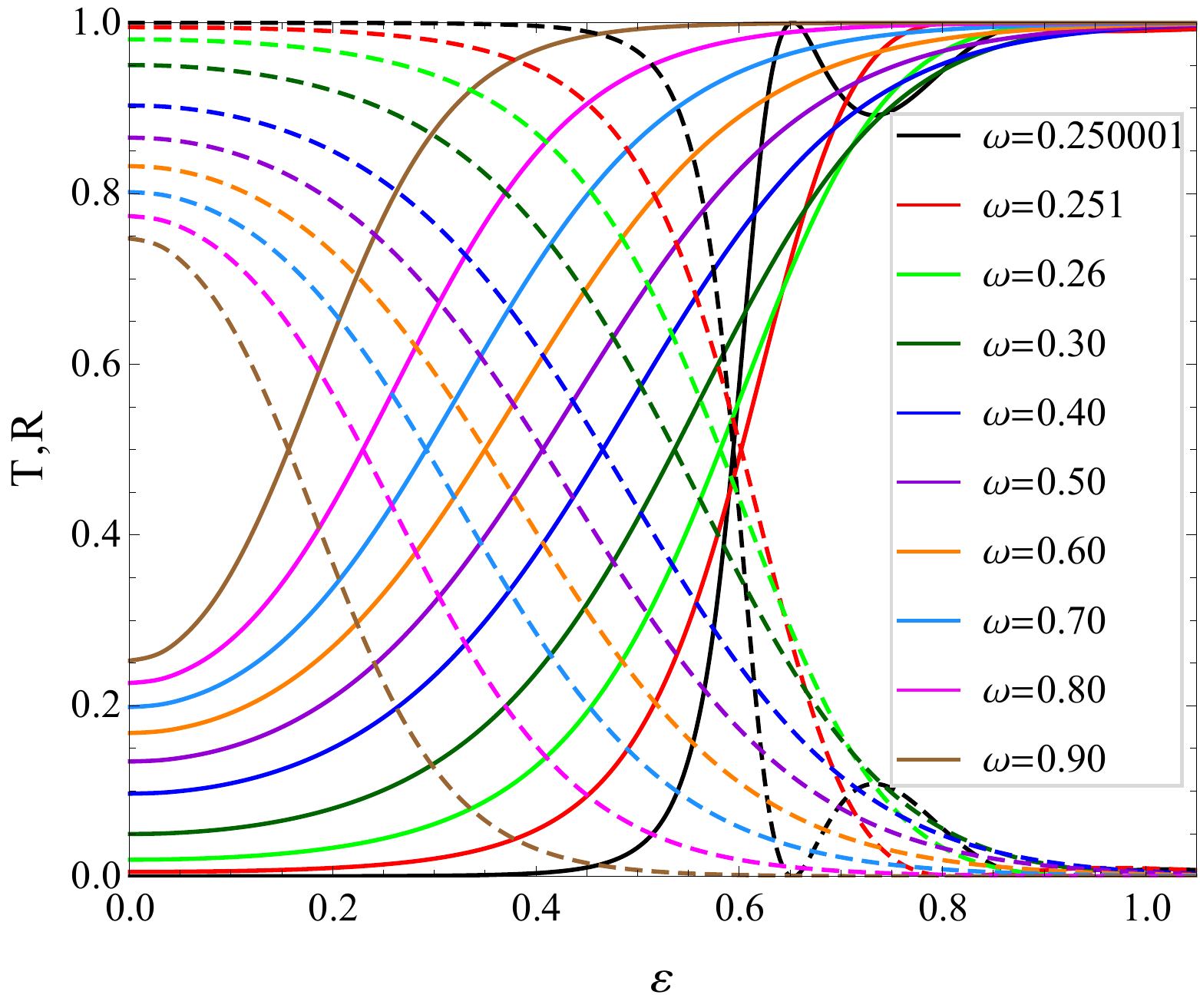}
\caption{Dependence of the transmission coefficient $T$ on the energy parameter
$\varepsilon$ for different values  of  the phase frequency of the Q-ball.  The
curves correspond to the parameters $h = 0.2$ and $G = 0.2$\hspace{1pt}.}
\label{fig:3}
\end{figure}

The behaviour of  the  curves  corresponding  to  the  elastic  matrix elements
$S_{1/2,\,1/2}$ and Eq.~(\ref{III:37}) tell us that the phase  shifts  $\delta(
\varepsilon)$ tend to zero for both $\varepsilon \rightarrow 0$ and $\varepsilon
\rightarrow \infty$.
In particular, it was  found  that the phase  shifts $\delta(\varepsilon)$ tend
to zero $\propto \varepsilon^{-1}$ as $\varepsilon \rightarrow \infty$.
Hence, the differences in the phase shifts $\Delta =\delta(0) - \delta(\infty)$
of the elastic matrix elements $S_{1/2,\,1/2}$ are equal to zero for all values
of the Yukawa coupling constant $G$.

\begin{figure}[tbp]\center
\includegraphics[width=0.5\textwidth]{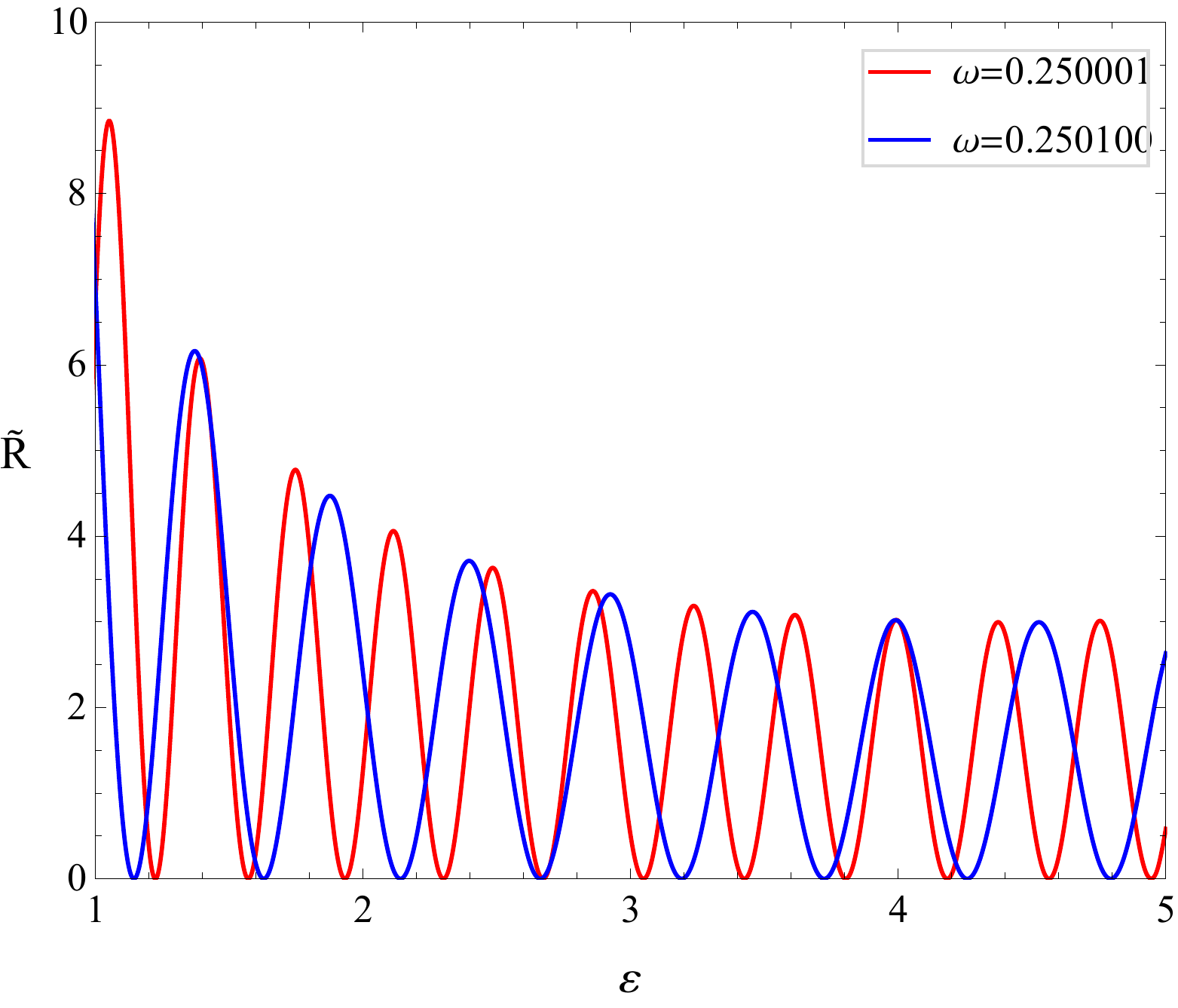}
\caption{Dependence  of  the  product  $\tilde{R}  = \exp(6.74  \varepsilon) R$
on the energy  parameter $\varepsilon$ for two values of the phase frequency of
the Q-ball.    The  curves  correspond  to  the  parameters  $h = 0.2$ and $G =
0.2$\hspace{1pt}.}
\label{fig:4}
\end{figure}

In contrast to $S_{1/2,\,1/2}$,  the  inelastic $S$-matrix elements $S_{-1/2,\,
1/2}$ are purely imaginary when  $\varepsilon = 0$;  they  tend  to  zero as $G
\rightarrow  0$ and to $-i$ as $G$ increases. 
As $\varepsilon$  increases,  the  matrix  elements  $S_{-1/2,\,1/2}$ leave the
imaginary axis and move along the dashed curves in Fig.~\ref{fig:2}.
For sufficiently large  $G$  and  sufficiently  small $\varepsilon$, the dashed
curves are close to the unitary circle.
With a further rise in  $\varepsilon$,  the  dashed  curves  tend to the origin,
indicating the exponential suppression of the fermion reflection as $\varepsilon
\rightarrow \infty$.

Next, we investigate the dependence  of  fermion-Q-ball scattering on the phase
frequency $\omega$  for fixed values of the other model parameters.
Figure~\ref{fig:3}  shows  the  dependence  of the transmission coefficient $T$
and  reflection  coefficient  $R$   on   the   energy  parameter $\varepsilon =
(\varepsilon_{+} + \varepsilon_{-})/2$ for  $h = 0.2$, $G = 0.2$, and different
values of the phase frequency $\omega$.
In the same way as in Fig.~\ref{fig:1}, we see that the curves $T(\varepsilon)$
and $R(\varepsilon)$ are even  functions  of  $\varepsilon$,  and  the values of
$T(0)$ ($R(0)$) are different from zero (unity).
However,  the  value  of  $T(0)$  ($R(0)$)   tends   to  zero  (unity)  in  the
thin-wall  regime  when  $\omega \rightarrow \omega_{\text{tn}} = 1/4$.
In Fig.~\ref{fig:3},  the   positions   of   the   intersection  points  of the
corresponding  curves  $T(\varepsilon)$  and  $R(\varepsilon)$  are  reasonably
well defined by Eqs.~(\ref{V:1}) and (\ref{V:2}), in which the parameter $G$ is
fixed and the variable is the phase frequency $\omega$.

A   characteristic  feature   of   the   curves   $R(\varepsilon, \omega)$  and
$T(\varepsilon,\omega)$ in Fig.~\ref{fig:3} is their resonance behaviour in the
thin-wall  regime as the phase frequency $\omega \rightarrow \omega_{\text{tn}}
= 1/4$.
Specifically, we can see that the  curve $R(\varepsilon, 0.250001)$ vanishes at
$\varepsilon \approx 0.655$.
Accordingly, the curve $T(\varepsilon, 0.250001)$ tends to unity at this point.
To better describe the resonance behaviour of the curves $R(\varepsilon,\omega)$
in the thin-wall regime,  Fig.~\ref{fig:4}  illustrates  the  dependence of the
product $\tilde{R}(\varepsilon, \omega) = \exp(6.74 \varepsilon) R(\varepsilon,
\omega)$ on the parameter $\varepsilon$.
The curves $\tilde{R}(\varepsilon, \omega)$  correspond  to  two  values of the
phase  frequency  $\omega$  in  the  close  vicinity  of  the  thin-wall  limit
$\omega_{\text{tn}} = 1/4$.
The exponential factor $\exp(6.74 \varepsilon)$ was selected empirically.
It is necessary to compensate for a decrease  in  $R(\varepsilon, \omega)$ with
an increase in $\varepsilon$.
It follows from Fig.~\ref{fig:4} that  there is  an infinite sequence of points
$\varepsilon_{k}$ such that $R(\varepsilon_{k}) = 0$.
As $k$ increases, the  difference  $\Delta \varepsilon_{k} = \varepsilon _{k} -
\varepsilon_{k-1}$ increases slightly and tends  to  a  constant  limit $\Delta
\varepsilon$.
This limit depends on  the  phase  frequency $\omega$, and decreases as $\omega
\rightarrow \omega_{\text{tn}}$.
Hence, the distance  between  the  successive  zeros  of  $R$ decreases with an
increase in the size of the Q-ball.
The successive zeros of $R$ are separated by peaks.
It follows from Fig.~\ref{fig:4}  that  the  height  of  these  peaks decreases
exponentially with an increase in $\varepsilon$.

The resonance behaviour of the curves shown in Figs.~\ref{fig:3} and \ref{fig:4}
can be explained at a qualitative level.
Due to its non-topological  nature, the one-dimensional Q-ball has two boundary
regions.
As $\omega \rightarrow \omega_{\text{tn}}$ (the  thin-wall regime), the spatial
size of the one-dimensional Q-ball increases,  representing a large homogeneous
central region bounded by two thin boundary regions.
The process of transmission (reflection)  of  a  fermion wave can occur at both
the left and right boundaries of the Q-ball. 
When the incident fermionic wave falls on the left boundary of the Q-ball, part
of it is reflected and the  remainder  passes  into the interior of the Q-ball.
Having reached the the right boundary  of  the Q-ball, part of the fermion wave
is transmitted and leaves the Q-ball, and  the  reminder  is reflected into the
Q-ball.
The reflected wave will then  reach  the left boundary of the Q-ball, where the
process is repeated.
Thus, the resulting reflected fermionic wave is determined by the superposition
of multiple reflected waves.
The zeros  of  $R$  correspond  to  destructive  wave interference, whereas its
peaks correspond to constructive interference.
The exponential decrease in the peak height of $R$ with increasing $\varepsilon$
occurs because the  amplitude  of  each  of  the reflected fermionic waves also
decreases exponentially as $\varepsilon$ increases.
Due to  unitarity  condition  (\ref{III:36}),  the  resonance  behaviour of the
reflection  coefficient  $R$  results   in   the   resonance  behaviour  of the
transmission coefficient $T$.
The situation is somewhat reminiscent of the passage of a light wave through an
antireflection lens.

\begin{figure}[tbp]\center
\includegraphics[width=0.5\textwidth]{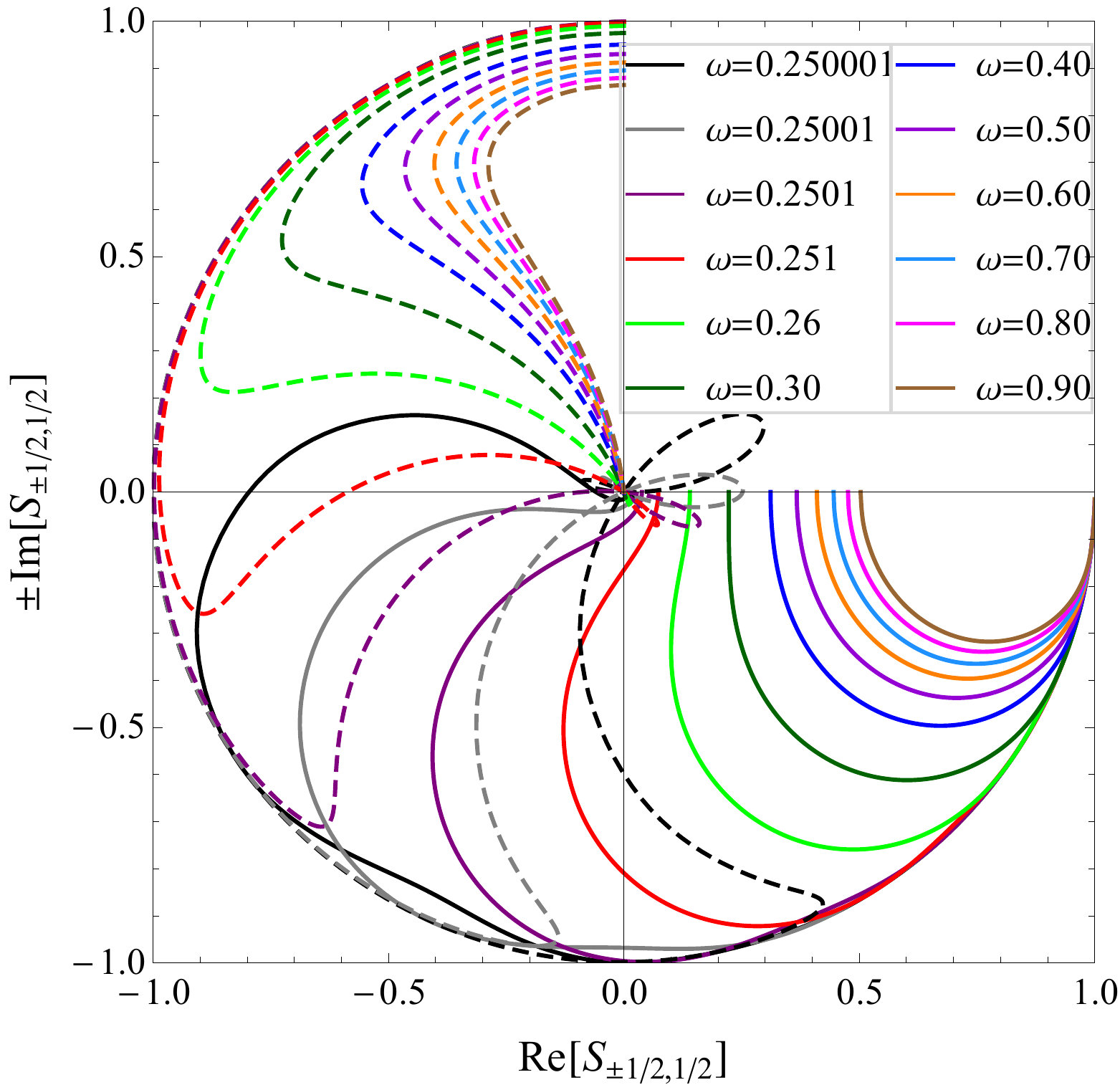}
\caption{Argand diagram  for  the  $S$-matrix  elements  $S_{1/2,\,1/2}$ (solid
curves) and complex conjugate $S$-matrix elements  $S^{*}_{-1/2,\,1/2}$ (dashed
curves) for different values  of  the phase frequency of the Q-ball. The curves
correspond to the parameters $h = 0.2$ and $G = 0.2$\hspace{1pt}.}
\label{fig:5}
\end{figure}

Figure~\ref{fig:5}  shows  an   Argand   diagram   for  the  elastic  $S$-matrix
elements $S_{1/2,\,1/2}$ and the complex conjugate inelastic $S$-matrix elements
$S^{*}_{-1/2,\,1/2}$ for different values of  the phase frequency $\omega$.
In contrast to  Fig.~\ref{fig:2}, the intersection  of  the  solid  and  dashed
curves cannot be avoided in this case.
In the same way as in  Fig.~\ref{fig:2}, the  solid curves start  ($\varepsilon
=0$) and end ($\varepsilon \rightarrow\infty$) on  the  real  axis, and tend to
unity as $\varepsilon \rightarrow \infty$.
This means that, as in  Fig.~\ref{fig:2},  the  difference  in the phase shifts
$\Delta=\delta(0)-\delta(\infty)$  of the elastic matrix elements $S_{1/2,1/2}$
is equal to zero  over  the  entire  allowable  range  of  the  phase frequency
$\omega$.

From Fig.~\ref{fig:5} it follows  that at $\varepsilon = 0$, the dashed curves,
which  correspond  to  the   complex   conjugate   inelastic   matrix  elements
$S_{-1/2,1/2}^{*}$, start on the imaginary axis.
As $\omega \rightarrow \omega_{\text{tn}}$ and $\varepsilon = 0$, the inelastic
matrix  elements tend to the point $(0, -i)$ lying on the unitary circle.
This corresponds to suppression of the elastic  channel in the thin-wall regime
at small  $\varepsilon$,  and  agrees  with  the  behaviour  of  the  curves in
Fig.~\ref{fig:3}.
As $\varepsilon$ increases, the  matrix  elements  leave the imaginary axis and
move along rather complex trajectories within the unitary circle.
The smaller the difference $\delta = \omega-\omega_{\text{tn}}$, the longer the
dashed curve in the vicinity of the unitary circle.
As $\varepsilon \rightarrow \infty$, the  dashed  curves  tend  to  the origin.
Furthermore, it follows from Fig.~\ref{fig:5} that for $\omega=0.251$, $0.2501$,
$0.25001$, and  $0.250001$,  the  dashed  curves  self-intersect at the origin.

\begin{figure}[tbp]\center
\includegraphics[width=0.5\textwidth]{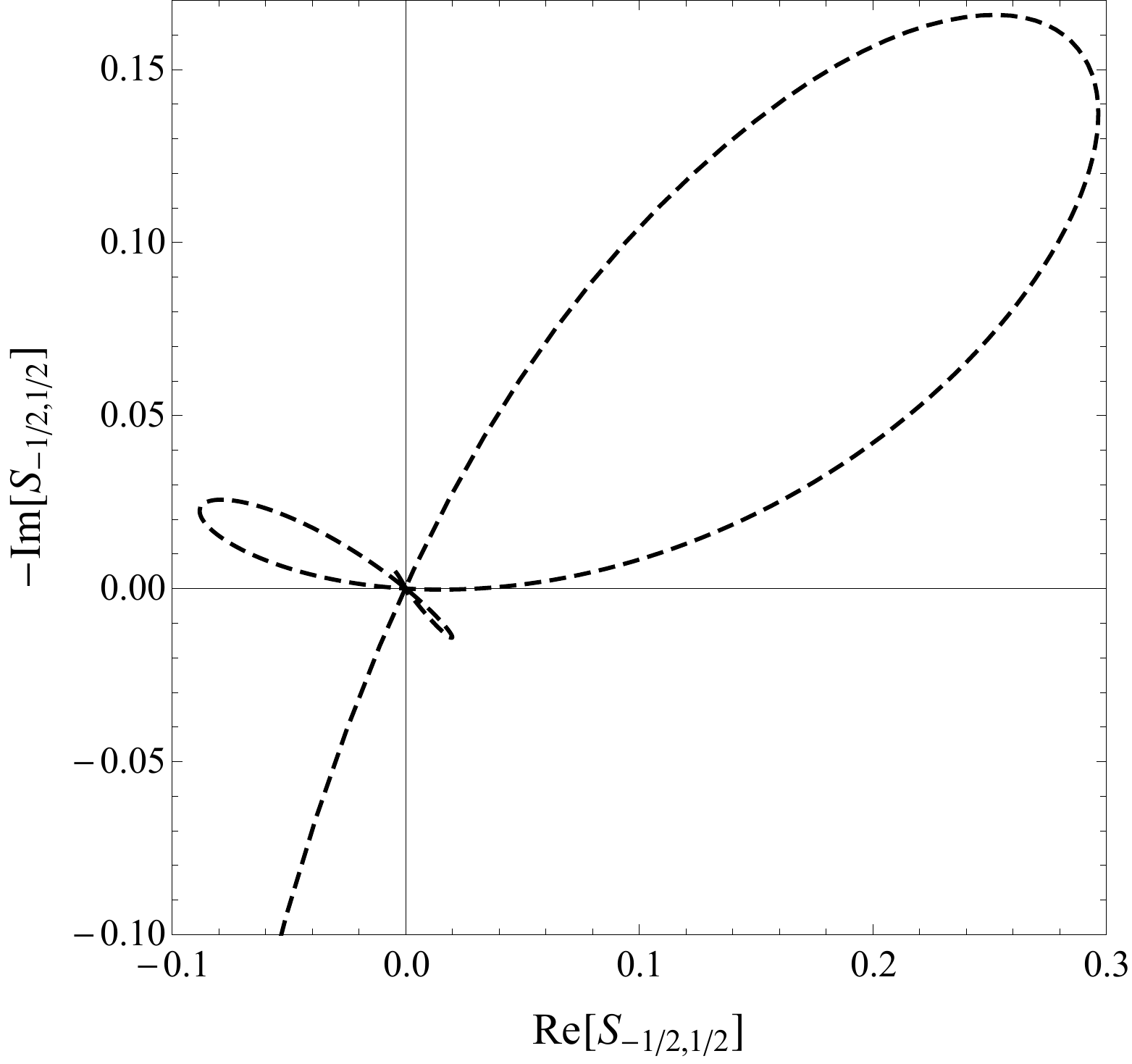}
\caption{Fragment of  an Argand  diagram  for  the complex conjugate $S$-matrix
element  $S^{*}_{-1/2,\,1/2}$   in  the  vicinity  of  the  origin.   The curve
corresponds  to   the   parameters   $h  =  0.2$,  $G  =  0.2$,  and  $\omega =
0.250001$\hspace{1pt}.}
\label{fig:6}
\end{figure}

For greater clarity, Fig.~\ref{fig:6}  shows  the behaviour of the dashed curve
corresponding to $\omega = 0.250001$ in the neighbourhood of the origin.
We can see that the  dashed  curve  crosses  the  origin  at  least four times.
In fact, from Fig.~\ref{fig:4} it  follows that the curve intersects the origin
an infinite number of times, as  the  intersections of the origin correspond to
the zeros of $R$ in Fig.~\ref{fig:4}.
These intersections, however, are indistinguishable  since in Fig.~\ref{fig:4},
the height of the peaks of  $R$  decreases  exponentially  with  an increase in
$\varepsilon$.
Note that according to  the unitarity condition (\ref{III:36}), the solid lines
touch the unitary circle  whenever  the  corresponding  dashed  lines cross the
origin.

\begin{figure}[tbp]\center
\includegraphics[width=0.5\textwidth]{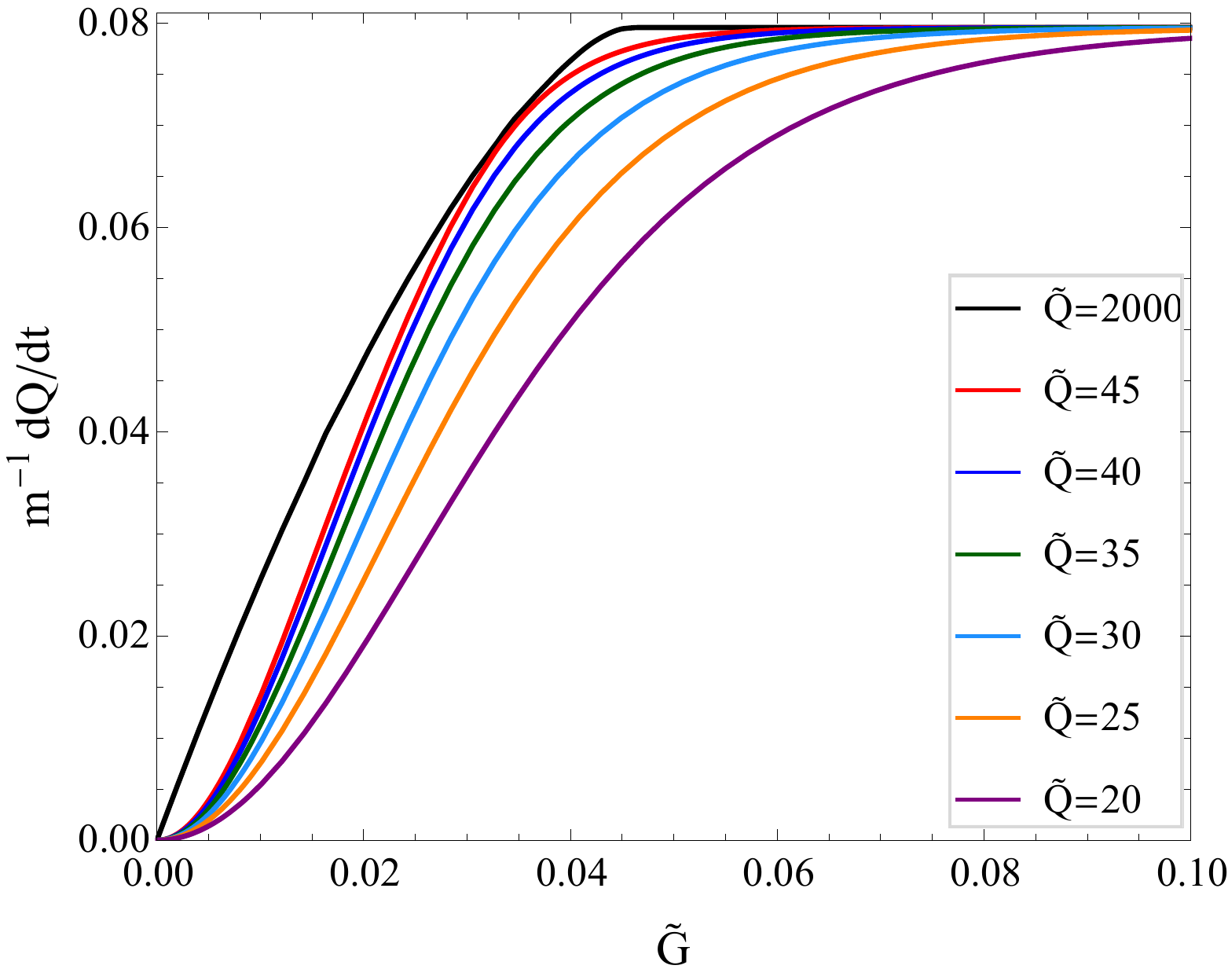}
\caption{Dependence of the evaporation rate  of  the  Noether charge $Q$ of the
Q-ball on the dimensionless Yukawa coupling constant  $\tilde{G}  =  g^{-1/2}G$
for different values  of the  reduced  Noether charge $\tilde{Q} = g m^{-2} Q$.
The curves correspond to the dimensionless parameter $\tilde{h}= m^{2}g^{-2}h =
0.2$\hspace{1pt}.}
\label{fig:7}
\end{figure}

In the following, we  present  numerical  results  for  the  evaporation of the
Q-ball.
As in Sec.~\ref{sec:IV},  when  discussing  this  subject,  we  use dimensional
quantities and denote  the  corresponding dimensionless analogues with a tilde.
Figure~\ref{fig:7} shows the dependence of the evaporation rate of the  Noether
charge $Q$  of  the  Q-ball  on  the  dimensionless  Yukawa  coupling  constant
$\tilde{G} = g^{-1/2}G$  for  different  values of the  reduced  Noether charge
$\tilde{Q} = g m^{-2}Q$.
Except for one case, the  curves  in  Fig.~\ref{fig:7}  correspond  to moderate
values of $\tilde{Q}$ at which the Q-ball begins to enter the thin-wall regime.
The remaining  curve  corresponds  to  a  significantly  greater  value  of the
reduced Noether charge $\tilde{Q} = 2 \times 10^{3}$.
At this value of $\tilde{Q}$, the $Q$-ball is in a pronounced thin-wall regime.
It was found  that  the  $m^{-1}dQ/dt$ curves  that  correspond  to  the larger
values of $\tilde{Q}$ are practically  indistinguishable  from  the black solid
curve in Fig.~\ref{fig:7}, which  can  therefore  be  regarded  as the limiting
curve.
From Fig.~\ref{fig:7}, it follows that as $\tilde{G}$ increases, all the $m^{-1}
dQ/dt$  curves  tend   to   the   limiting   value  $m^{-1} \left. dQ/dt \right
\vert_{\max}  \approx  \tilde{\omega}_{\text{tn}}/\pi   =  0.0796$  defined  by
Eq.~(\ref{IV:3}).
At the same time,  as  the  reduced Noether charge  $\tilde{Q}$  increases, the
$m^{-1}dQ/dt$ curves tend to the limiting curve in Fig.~\ref{fig:7}.

\begin{figure}[tbp]\center
\includegraphics[width=0.5\textwidth]{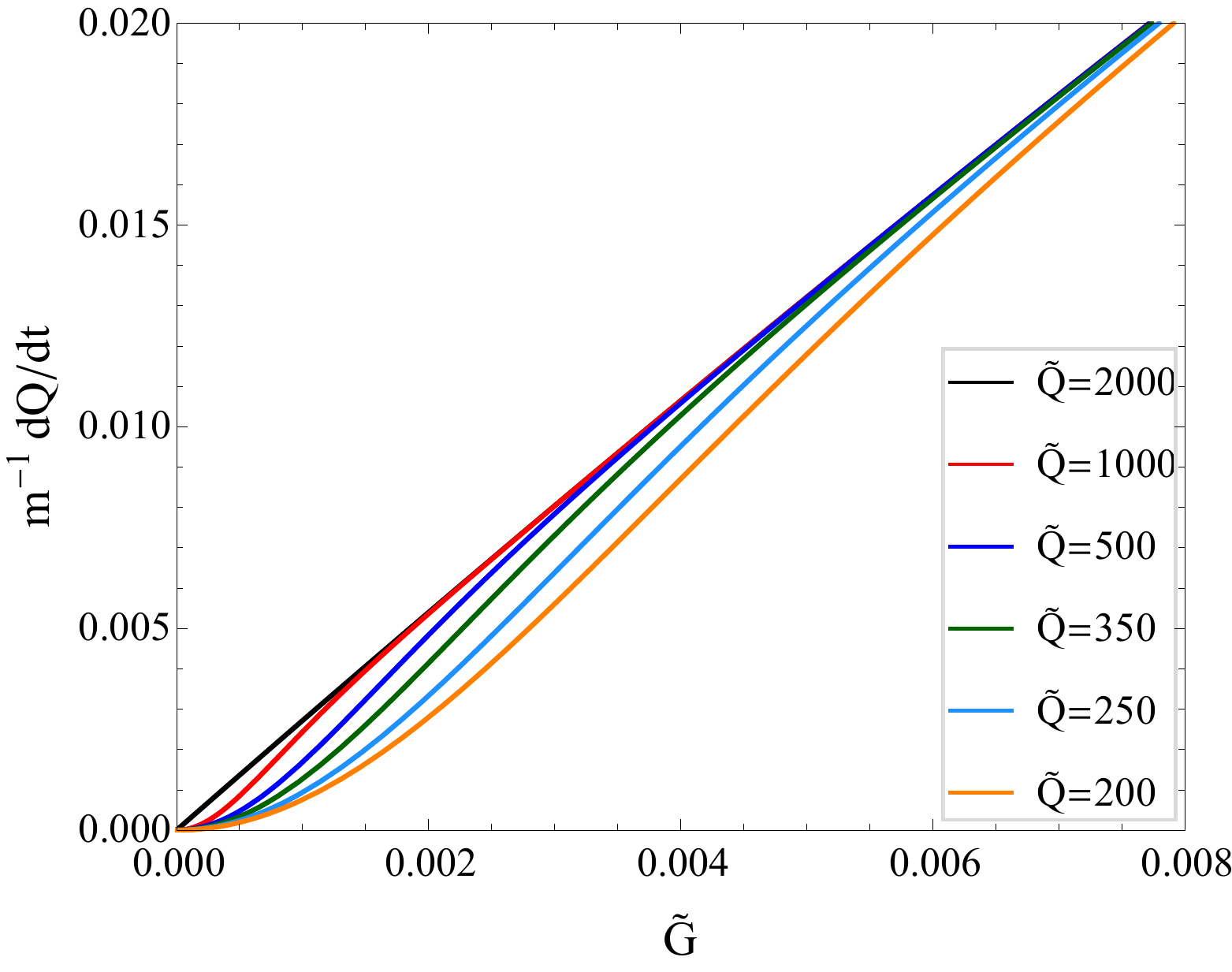}
\caption{Dependence of the evaporation  rate  of the  Noether charge $Q$ of the
Q-ball on the dimensionless Yukawa coupling constant  $\tilde{G}  =  g^{-1/2}G$
for different values  of  the  reduced Noether charge $\tilde{Q} = g m^{-2} Q$.
The curves correspond to the dimensionless  parameter $\tilde{h} = m^{2}g^{-2}h
= 0.2$\hspace{1pt}.}
\label{fig:8}
\end{figure}

To better understand the  limiting   behaviour   of   the $m^{-1}dQ/dt$ curves,
Fig.~\ref{fig:8} shows these curves in the vicinity of the origin.
Compared  to  Fig.~\ref{fig:7}, the  curves  in  Fig.~\ref{fig:8} correspond to
larger $\tilde{Q}$, and thus their limiting behaviour is more pronounced.
Using the thin-wall  approximation,  in  which  the  $x$-dependent  part of the
profile functions in Eqs.~(\ref{II:11a}) and (\ref{II:11b}) is  replaced   by a
rectangular function, it can be shown that as $G\rightarrow 0$, the evaporation
rate
\begin{equation}
dQ/dt\sim 8G^{2}g^{-1}\left(m^{2}-\omega_{\text{tn}}^{2}\right) L,  \label{V:3}
\end{equation}
where the linear size $L$ of the Q-ball is related to its Noether charge $Q$ by
Eq.~(\ref{IV:6}).
It follows from Eq.~(\ref{V:3}) that  the  first  derivative of the evaporation
rate $dQ/dt$ with respect to $G$ vanishes at $G = 0$.
The second  derivative, however, is  proportional  to  $L \approx Q/j_{0}$, and
hence it increases indefinitely in the  thin-wall regime  when both $L$ and $Q$
tend to infinity.

It follows from  Fig.~\ref{fig:8}  that  as $\tilde{Q} \rightarrow \infty$, the
$m^{-1}dQ/dt$ curves degenerate into a limiting  straight  line in the vicinity
of the origin.
Using the thin-wall approximation, it can be  shown that this limiting straight
line is described by the expression
\begin{equation}
dQ/dt  \approx  \pi^{-1}G \left\vert \phi \left(0\right)\right\vert
= 2^{3/2}\pi^{-1}G g^{-1/2}\left(m^{2}-\omega_{\text{tn}}^{2}\right)^{1/2}.
                                                                    \label{V:4}
\end{equation}
From Figs.~\ref{fig:7} and \ref{fig:8}, it follows  that for $\tilde{Q} \gtrsim
200$, the shape of the $m^{-1}dQ/dt$ curves practically ceases to depend on the
magnitude of $\tilde{Q}$,  and  the  reduced  evaporation rate $m^{-1}dQ/dt$ is
determined only by the  value  of  the  dimensionless  Yukawa coupling constant
$\tilde{G} = g^{-1/2}G$.
In this case, the Noether charge  of  the  Q-ball decreases linearly with time:
\begin{equation}
Q\left(t\right) \approx Q\left(0\right)-m \Gamma( \tilde{G})t,      \label{V:5}
\end{equation}
where  $\Gamma( \tilde{G})$   is  the  limiting  $m^{-1}dQ/dt$  curve  shown in
Figs.~\ref{fig:7} and \ref{fig:8}.
It follows that the  evaporation  of  the  Noether  charge  results in a finite
lifetime  of  the  Q-ball, $\tau  \approx  Q/(m  \Gamma(\tilde{G}))  \ge  \pi Q
\omega_{\text{tn}}^{-1}$.
We see that the lifetime of the Q-ball $\tau$ is $\propto Q$, and can therefore
be arbitrarily large in the thin-wall regime since $\underset{\omega\rightarrow
\omega_{\text{tn}}}{\lim}Q\left(\omega\right) = \infty$.
This is possible because in the thin-wall regime, the evaporation of the Q-ball
occurs only at its boundaries rather than within its interior.
In the latter case, the decrease in the charge would be exponential rather than
linear, meaning  that  the  lifetime  of  the  Q-ball  would  not depend on the
magnitude of its charge.

\section{Conclusion}                                            \label{sec:VI}

In  the  present  paper,  we study  the  scattering of massless fermions in the
background field of a one-dimensional Q-ball.
Since the one-dimensional Q-ball solution is known in  analytical form, we were
able to obtain analytical expressions for the fermionic wave functions in terms
of the local Heun functions.
The analytical form of the fermionic wave functions makes it possible to derive
general analytical expressions for  the transmission coefficient $T$, reflection
coefficient $R$, and corresponding $S$-matrix elements.
In turn, the existence  of  these  general  expressions  greatly  simplifies the
numerical study of the properties of fermion-Q-ball scattering.

The main feature of fermion-Q-ball scattering is the resonance behaviour of the
transmission and reflection coefficients in the thin-wall regime.
This resonance behaviour consists  of  the existence of an infinite sequence of
values $\varepsilon_{i}$ of the energy parameter, such that $R(\varepsilon_{i})
= 0$ and $T(\varepsilon_{i}) = 1$.
The zeros of the  reflection  coefficient  $R$  are  separated  by  peaks whose
height decreases exponentially with an increase in their sequence number.
The reason for the resonance behaviour of  the  coefficients $R$ and $T$ is due
to the nontopological nature  of  the  Q-ball,  resulting  in  the existence of
two boundary regions for the one-dimensional Q-ball.
The existence of these  two  boundaries  makes possible multiple reflections of
fermionic waves inside the Q-ball.
The  resonance  structure  of  the  coefficient  $R$  ($T$)  results  from  the
interference of these  multiple reflected waves at the left (right) boundary of
the Q-ball.

Unlike a  one-dimensional  Q-ball,  a  kink  is  a  one-dimensional topological
soliton.
The kink interpolates between two topologically different vacua, and essentially
consists of a single transition region. 
This  makes  the  multiple  reflection  of  waves  impossible  in  fermion-kink
scattering.
As  a  result,  the  energy  dependence  of  the  transmission  and  reflection
coefficients has no resonance structure in this case.

The Q-ball solution is  $\propto \exp(-i \omega t)$, and hence has a nontrivial
time dependence.
As a result, the energy of fermions is not conserved when they are reflected in
the background field of the Q-ball.
Moreover, when the energy parameter $\varepsilon \in (-\omega/2, \omega/2)$, it
becomes possible to  produce  fermion-antifermion  pairs  that  carry  away the
energy and Noether charge of the Q-ball \cite{ccgm, mv, clark}.
In the thin-wall regime,  the  Noether  charge,  energy, and linear size of the
Q-ball  become   large,   and   the  one-dimensional  Q-ball  has  a  spatially
homogeneous distribution  of   energy  and  charge except for two thin boundary
regions.
In this case,  pair production  cannot occur in the interior of the Q-ball, and
is possible only at its boundaries.
Pair production results in the evaporation  of the Noether charge, leading to a
decrease in the linear size of the Q-ball.

In the leading order of the semi-classical  approximation, the evaporation rate
of the Noether  charge  can  be  expressed  in  terms  of  the  integral of the
reflection  coefficient  $R$  over  the  region  $\varepsilon  \in  (-\omega/2,
\omega/2)$.
In our  case,  the  condition  for  the  applicability  of  the  semi-classical
approximation is the smallness  of  the  dimensionless combination $gm^{-2}$ at
fixed values of the other dimensionless combinations.
We studied the dependence of the evaporation rate $dQ/dt$ on  the dimensionless
Yukawa coupling constant  $\tilde{G} = g^{-1/2} G$  for  a  number of values of
the reduced Noether charge $\tilde{Q} = g m^{-2} Q$, and found that the $m^{-1}
dQ/dt$ curves tend  to  a  limiting  curve  as  $\tilde{Q} \rightarrow \infty$.
As $\tilde{G}$ increases,  this  limiting  curve  tends  to  the limiting value
$\omega_{\text{tn}}/(\pi m)$,   which   depends   only   on  the  dimensionless
combination $\tilde{h} = m^{2}g^{-2}h$.

It follows from the results  presented  in  this  work  that  in  the thin-wall
regime, the character of  the  fermion-Q-ball  interaction is determined by two
parameters: the phase frequency $\omega$ and  the parameter $\varepsilon_{1/2}$
defined in Eq.~(\ref{V:2}).
There  are  two  possible cases: $\omega/2 < \varepsilon_{1/2}$ and $\omega/2 >
\varepsilon_{1/2}$.
In the first case, if the energy parameter $\varepsilon\in(-\omega/2,\omega/2)$,
then  evaporation   of   the   Q-ball  takes  place, if $\omega/2 < \varepsilon
\lesssim \varepsilon_{1/2}$, then  almost  all  of  the incident  fermions  are
reflected from the Q-ball, and if $\varepsilon \gtrsim \varepsilon_{1/2}$, then
almost all of the incident  fermions pass through the Q-ball.
In the second case, if the energy parameter $\varepsilon\in(-\omega/2,\omega/2)
$, then evaporation  of  the  Q-ball  takes  place as in the first case, and if
$\varepsilon > \omega/2$, then almost all of the incident fermions pass through
the Q-ball.
We see that if $\omega/2  >  \varepsilon_{1/2}$, then reflection of fermions is
practically absent.

A characteristic property  of  fermion-Q-ball  scattering is that the parameter
$\varepsilon_{1/2}$ is approximately  equal  to  the effective mass $m_{\psi\!,
\text{eff}}$ of the fermion in the background field of the Q-ball.
In turn, the effective mass $m_{\psi\!,\text{eff}}\approx 3^{1/2}2^{-1}\tilde{G}
\tilde{h}^{-1/2} m$.
In the semi-classical regime, the dimensionless coupling constants $\tilde{G} =
g^{-1/2}G$ and $\tilde{h} = m^{2} g^{-2} h$ are $\lesssim 1$.
It follows that in this regime, the effective fermion mass $m_{\psi\!,\text{eff
}} \lesssim m$, where $m$ is the mass of the scalar boson.
We see that  in  the  semi-classical  regime,  the  mass  of  the  scalar boson
significantly affects the character of fermion-Q-ball scattering.





\bibliographystyle{elsarticle-num}
\bibliography{article}

\begin{thebibliography}{10}
\expandafter\ifx\csname url\endcsname\relax
  \def\url#1{\texttt{#1}}\fi
\expandafter\ifx\csname urlprefix\endcsname\relax\def\urlprefix{URL }\fi
\expandafter\ifx\csname href\endcsname\relax
  \def\href#1#2{#2} \def\path#1{#1}\fi

\bibitem{lee_pang_1992}
T.~D. Lee, Y.~Pang, Phys. Rep. 221 (1992) 251.

\bibitem{radu_volkov_2008}
E.~Radu, M.~Volkov, Phys. Rep. 468 (2008) 101.

\bibitem{Manton}
N.~Manton, P.~Sutclffe, Topological Solitons, Cambridge University Press,
  Cambridge, 2004.

\bibitem{Shnir}
Y.~M. Shnir, Topological and Non-Topological Solitons in Scalar Field Theories,
  Cambridge University Press, Cambridge, 2018.

\bibitem{rosen_1968_a}
G.~Rosen, J. Math. Phys. (N.Y.) 9 (1968) 996.

\bibitem{coleman_1985}
S.~Coleman, Nucl. Phys. B 262 (1985) 263.

\bibitem{saf1}
A.~Safian, S.~Coleman, M.~Axenides, Nucl. Phys. B 297 (1988) 498.

\bibitem{saf2}
A.~Safian, Nucl. Phys. B 304 (1988) 403.

\bibitem{rosen_1968_b}
G.~Rosen, J. Math. Phys. (N.Y.) 9 (1968) 999.

\bibitem{klee}
K.~Lee, J.~A. Stein-Schabes, R.~Watkins, L.~M. Widrow, Phys. Rev. D 39 (1989)
  1665.

\bibitem{lee_yoon}
C.~H. Lee, S.~U.Yoon, Mod. Phys. Lett. A 6 (1991) 1479.

\bibitem{anag}
K.~N. Anagnostopoulos, M.~Axenides, E.~G. Floratos, N.~Tetradis, Phys. Rev. D
  64 (2001) 125006.

\bibitem{levi}
T.~S. Levi, M.~Gleiser, Phys. Rev. D 66 (2002) 087701.

\bibitem{ardoz_2009}
H.~Arodz, J.~Lis, Phys. Rev. D 79 (2009) 045002.

\bibitem{benci}
V.~Benci, D.~Fortunato, J. Math. Phys. (N.Y.) 52 (2011) 093701.

\bibitem{tamaki_2014}
T.~Tamaki, N.~Sakai, Phys. Rev. D 90 (2014) 085022.

\bibitem{gulamov_2014}
I.~E. Gulamov, E.~Y. Nugaev, M.~N. Smolyakov, Phys. Rev. D 89 (2014) 085006.

\bibitem{brihaye_2014}
Y.~Brihaye, V.~Diemer, B.~Hartmann, Phys. Rev. D 89 (2014) 084048.

\bibitem{hong_2015}
J.~Hong, Y.~Kim, P.~Y. Pac, Phys. Rev. Lett. 64 (1990) 2230.

\bibitem{gulamov_2015}
I.~E. Gulamov, E.~Y. Nugaev, A.~G. Panin, M.~N. Smolyakov, Phys. Rev. D 92
  (2015) 045011.

\bibitem{loginov_prd_102}
A.~{\relax Yu}. Loginov, V.~V. Gauzshtein, Phys. Rev. D 102 (2020) 025010.

\bibitem{kus_plb_1997_405}
A.~Kusenko, Phys. Lett. B 405 (1997) 108.

\bibitem{kst_1998}
A.~Kusenko, M.~Shaposhnikov, P.~Tinyakov, Pis'ma Zh. Exp. Teor. Fiz. 67 (1998)
  229.

\bibitem{enqmcd_1998}
K.~Enqvist, J.~McDonald, Phys. Lett. B 425 (1998) 309.

\bibitem{dksh_plb_417}
G.~Dvali, A.~Kusenko, M.~Shaposhnikov, Phys. Lett. B 417 (1998) 99.

\bibitem{kusshp_plb_418}
A.~Kusenko, M.~Shaposhnikov, Phys. Lett. B 418 (1998) 46.

\bibitem{enqmcd_1999}
K.~Enqvist, J.~McDonald, Nucl. Phys. B 538 (1999) 321.

\bibitem{kk_2000}
S.~Kasuya, M.~Kawasaki, Phys. Rev. D 61 (2000) 041301.

\bibitem{enqmzm_2003}
K.~Enquist, A.~Mazumdar, Phys. Rep. 380 (2003) 99.

\bibitem{kky_2013a}
A.~Kamada, M.~Kawasaki, M.~Yamada, Phys. Lett. B 719 (2013) 9.

\bibitem{kky_2013b}
S.~Kasuya, M.~Kawasaki, M.~Yamada, Phys. Lett. B 726 (2013) 1.

\bibitem{cotner_2017a}
E.~Cotner, A.~Kusenko, Phys. Rev. Lett. 119 (2017) 031103.

\bibitem{cotner_2017b}
E.~Cotner, A.~Kusenko, Phys. Rev. D 96 (2017) 103002.

\bibitem{ccgm}
A.~G. Cohen, S.~R. Coleman, H.~Georgi, A.~Manohar, Nucl. Phys. B 272 (1986)
  301.

\bibitem{kvng}
A.~V. Kovtun, E.~Y. Nugaev, Mod. Phys. Lett. A 32 (2017) 1750198.

\bibitem{mv}
T.~Multamaki, I.~Vilja, Nucl. Phys. B 574 (2000) 130.

\bibitem{clark}
S.~S. Clark, Nucl. Phys. B 756 (2006) 38.

\bibitem{kus_plb_1997_404}
A.~Kusenko, Phys. Lett. B 404 (1997) 285.

\bibitem{paccetti}
F.~P. Correia, M.~Schmidt, Eur. Phys. J. C 21 (2001) 181.

\bibitem{Ronveaux}
A.~Ronveaux (Ed.), Heun's differential equations, Oxford University Press,
  Oxford, 1995.

\bibitem{DLMF}
F.~W.~J. Olver, D.~W. Lozier, R.~F. Boisvert, C.~W. Clark (Eds.), NIST Handbook
  of Mathematical Functions, Cambridge University Press, Cambridge, 2010.

\bibitem{LandauIII}
L.~D. Landau, E.~M. Lifshitz, Quantum Mechanics: Non-Relativistic Theory. Vol.
  3 (3rd ed.), Pergamon Press, Oxford, 1977.

\bibitem{Goldberger}
M.~L. Goldberger, K.~M. Watson, Collision Theory, John Wiley \& Sons, New York,
  1967.

\bibitem{Taylor}
J.~R. Taylor, Scattering Theory: Quantum Theory on Nonrelativistic Collisions,
  John Wiley \& Sons, New York, 1972.

\end{thebibliography}






\end{document}